\documentclass[acmlarge]{acmart}

\AtBeginDocument{%
  \providecommand\BibTeX{{%
    \normalfont B\kern-0.5em{\scshape i\kern-0.25em b}\kern-0.8em\TeX}}}

\usepackage{graphicx}
\usepackage[disable]{todonotes}
\usepackage{hyperref}
\usepackage{caption}
\usepackage{subcaption}
\usepackage{multirow}
\usepackage{multicol}
\usepackage{acronym}
\usepackage{soul}

\setcopyright{rightsretained}

\begin{document}

\acmJournal{DTRAP}
\acmYear{2023} \acmVolume{1} \acmNumber{1} \acmArticle{1} \acmMonth{1} \acmPrice{}\acmDOI{10.1145/3617897}
\title{Machine Learning (In) Security: A Stream of Problems}


\author{Fabrício Ceschin}
\orcid{0000-0001-6853-8083 }
\email{fjoceschin@inf.ufpr.br}
\affiliation{%
  \institution{Federal University of Paraná}
  \streetaddress{Rua Cel. Francisco H. dos Santos, 100}
  \city{Curitiba}
  \state{PR}
  \postcode{81531-980}
  \country{Brazil}}
\affiliation{%
  \institution{Georgia Institute of Technology}
  \streetaddress{756 West Peachtree Street NW}
  \city{Atlanta}
  \state{GA}
  \postcode{30308-4016}
  \country{USA}
}
\author{Marcus Botacin}
\orcid{0000-0001-6870-1178}
\email{botacin@tamu.edu}
\affiliation{%
  \institution{Texas A\&M University}
  \department{Department of Computer Science \& Engineering}
  \city{College Station}
  \state{TX}
  \postcode{77843-3127}
  \country{USA}
}
\author{Albert Bifet}
\email{abifet@waikato.ac.nz}
\affiliation{%
  \institution{University of Waikato}
  \department{Department of Computer Science}
  \city{Waikato}
  \state{Hamilton}
  \postcode{}
  \country{New Zealand}
}
\author{Bernhard Pfahringer}
\email{bernhard@waikato.ac.nz}
\affiliation{%
  \institution{University of Waikato}
  \department{Department of Computer Science}
  \city{Waikato}
  \state{Hamilton}
  \postcode{}
  \country{New Zealand}
}
\author{Luiz S. Oliveira}
\email{lesoliveira@inf.ufpr.br}
\affiliation{%
  \institution{Federal University of Paraná}
  \streetaddress{Rua Cel. Francisco H. dos Santos, 100}
  \city{Curitiba}
  \state{PR}
  \postcode{81531-980}
  \country{Brazil}
}
\author{Heitor Murilo Gomes}
\email{heitor.gomes@vuw.ac.nz}
\affiliation{%
  \institution{Victoria University of Wellington}
  \department{School of Engineering and Computer Science}
  \city{Wellington}
  \state{}
  \postcode{}
  \country{New Zealand}
}
\author{André Grégio}
\email{gregio@inf.ufpr.br}
\affiliation{%
  \institution{Federal University of Paraná}
  \streetaddress{Rua Cel. Francisco H. dos Santos, 100}
  \city{Curitiba}
  \state{PR}
  \postcode{81531-980}
  \country{Brazil}
}

\definecolor{darkgreen}{RGB}{62, 199, 47}
\newcommand{\tiago}[1]{\textcolor{darkgreen}{#1}}
\newcommand{\fabricio}[1]{\todo[inline, color=darkgreen]{\textbf{Fabricio:} #1}}
\newcommand{\gregio}[1]{\todo[inline, color=blue]{Gregio: #1}}
\newcommand{\luiz}[1]{\todo[inline, color=orange]{Luiz: #1}}
\newcommand{\heitor}[1]{\todo[inline, color=purple]{Heitor: #1}}
\newcommand{\marcus}[1]{\todo[inline, color=gray]{Marcus: #1}}
\newcommand{\saman}[1]{\todo[inline, color=yellow]{Saman: #1}}
\newcommand{\review}[1]{#1}
\newcommand{\reviewdtrap}[1]{\textcolor{black}{#1}}
\newcommand{\improve}[1]{\textcolor{black}{#1}}
\newcommand{\blind}[0]{\textcolor{black}{\url{https://secret.inf.ufpr.br/machine-learning-in-security-checklist/}}}

\newcommand{\fabricioreview}[2]{
    \todo[inline, color=orange]{\textbf{Reviewer:} #1}
    \todo[inline, color=darkgreen]{\textbf{Fabricio:} #2}

}

\newacro{IDS}{Intrusion Detection System}
\newacro{ML}{Machine Learning}
\newacro{NIDS}{Network Intrusion Detection Systems}
\newacro{HIDS}{Host-based Intrusion Detection Systems}
\newacro{OS}{Operating System}
\newacro{STIDE}{Sequence Time-Delay Embedding}
\newacro{HMM}{Hidden Markov Model}
\newacro{CA}{Continuous Authentication}

\renewcommand{\shortauthors}{F. Ceschin et al.}

\begin{abstract}
    Machine Learning (ML) has been widely applied to cybersecurity and is considered state-of-the-art for solving many of the open issues in that field. However, it is very difficult to evaluate how good the produced solutions are, since the challenges faced in security may not appear in other areas. One of these challenges is the concept drift, which increases the existing arms race between attackers and defenders: malicious actors can always create novel threats to overcome the defense solutions, which may not consider them in some approaches. Due to this, it is essential to know how to properly build and evaluate an ML-based security solution. In this paper, we identify, detail, and discuss the main challenges in the correct application of ML techniques to cybersecurity data. We evaluate how concept drift, evolution, delayed labels, and adversarial ML impact the existing solutions. Moreover, we address how issues related to data collection affect the quality of the results presented in the security literature, showing that new strategies are needed to improve current solutions. Finally, we present how existing solutions may fail under certain circumstances, and propose mitigations to them, presenting a novel checklist to help the development of future ML solutions for cybersecurity.
\end{abstract}

\ccsdesc[500]{Security and privacy~Intrusion/anomaly detection and malware mitigation}
\ccsdesc[500]{Security and privacy}
\ccsdesc[500]{Security and privacy~Systems security}


\keywords{machine learning, cybersecurity, data streams}

\maketitle

\section{Introduction}
\label{sec:introduction}

The massive amount of data produced daily demands automated solutions capable of keeping \acf{ML} models updated and working properly, even with all emerging threats that constantly try to evade these models. This arms race between attackers and defenders moves the cybersecurity research forward: malicious actors are continuously creating new variants of attacks, exploring new vulnerabilities, and crafting adversarial samples, whereas 
\reviewdtrap{researchers}
are trying to counter those threats and improve detection models. For instance, 68\% of phishing emails blocked by GMail are different from day to day~\cite{elie19}, requiring Google to update and adapt its security components regularly. \reviewdtrap{In addition, 30\% of Windows malware released daily may belong to unknown families, which requires ML models to be updated with behavioral reports to detect drifting samples~\cite{10.1145/3291061}. Furthermore, at least 53\% of organizations are deploying AI in different areas of cybersecurity~\cite{9568267}.}
Thus, applying \ac{ML} on cybersecurity is a challenging endeavor. One of the main challenges is the volatility of the data used for building models as attackers constantly develop adversarial samples to avoid detection\reviewdtrap{~\cite{10.1145/3375894.3375898, 10.1145/3433667.3433669, apruzzese2023realgradients}}. This leads to a situation where the models need to be constantly updated to keep track of new attacks. 
Another challenge is related to the application of fully supervised methods since the class labels tend to depict an extreme imbalance, i.e., dozens of attacks \reviewdtrap{among} thousands of benign samples.
Labeling such instances is also problematic as it requires domain knowledge and it can detain the learning method, i.e. the analyst labeling the data is a bottleneck in the learning process. 
This motivates the development of semi-supervised and anomaly detection methods~\cite{9833671, Apruzzese_2022}.

To improve the process of continuously updating an \ac{ML} cybersecurity solution, the adoption of stream learning (also incremental learning or online learning) algorithms are recommended so they can operate in real-time using a reasonable amount of resources, considering that we have limited time and memory to process each new sample incrementally, predict samples at any time, and adapt to changes~\cite{MOA-Book-2018}. 
However, many works in the literature do not consider these challenges when proposing a solution, which makes them not feasible in reality. \reviewdtrap{Thus, multiple cybersecurity research papers cannot be straightforwardly applied to solve real-world problems and sometimes are not focused on ``machine learning that matters''~\cite{DBLP:journals/corr/abs-1206-4656}, i.e., there is a high discrepancy between research and practice~\cite{Apruzzese_2023}.
}

Some previous work reported the relevance of some of these problems and provided research directions. 
Ede et al. studied how to automatically correlate security events and automate parts of the security operator workload in a semi-supervised manner~\cite{9833671}.
Papernot et al. systematized findings on ML security and privacy, focusing on attacks identified on \ac{ML} systems and defense solutions, creating a threat model for \ac{ML}, and categorizing attacks and defenses within an adversarial framework~\cite{8406613}.
Maiorca et al. explored adversarial attacks against PDF (Portable Document Format) malware detectors, highlighting how the arms race between attackers and defenders has evolved over the last decade~\cite{maiorca2019towards}. 
%
Arnaldo et al. described a set of challenges faced when developing a real cybersecurity \ac{ML} platform, stating that many research papers are not valid in many use cases, with a special focus on label acquisition and model deployment~\cite{arnaldo2019holy}. 
Kaur et al. presented a comparative analysis of some approaches used to deal with imbalanced data (pre-processing methods, algorithmic-centered techniques, and hybrid ones), applying them to different data distributions and application areas~\cite{kaur2019systematic}. Gibert et al. listed a set of methods and features used in a traditional \ac{ML} workflow for malware detection and classification in the literature with emphasis on deep learning approaches, exploring some of their limitations and challenges, such as class imbalance, open benchmarks, concept drift, adversarial learning and interpretability of the models~\cite{GIBERT2020102526}. 

\reviewdtrap{Boenisch et al. investigated the security and privacy awareness of ML practitioners through an online survey and concluded that their awareness of threats and ML security practices is relatively low, as well as their familiarity with security and privacy libraries for ML~\cite{10.1145/3473856.3473869}, which evidence that this kind of study is very important for the community.} 
\reviewdtrap{Grosse et al. found that there are occurrences of adversarial ML attacks but regular security threats still pose a larger concern in industry~\cite{Grosse_2023}. The authors deduce that adversarial machine learning in practice is not as common as regular security problems, but that monitoring ML security might be beneficial.}
\reviewdtrap{Bieringer et al. showed that practitioners confuse ML security with threats that are not directly related to ML and consider that the security of ML is related to the entire workflow that consists of multiple components~\cite{281210}.} 
\reviewdtrap{Apruzzese et al. highlight the lists of all the tasks where ML outperforms traditional security mechanisms and reveal some of the challenges that require the contribution of all stakeholders involved to improve the quality of ML-based security solutions~\cite{Apruzzese_2023}.} Arp et al. conducted a study of 30 papers from top-tier security conferences within the past 10 years and showed that many pitfalls are widespread in the current security literature~\cite{arpEtAl2022}. \reviewdtrap{Finally, Apruzzese et al. state positions that increase the real-world impact of future research, bringing researchers and practitioners closer to improving the security of ML-based solutions~\cite{apruzzese2023realgradients}.}


In this work, we present a broad collection of gaps, pitfalls, and challenges that are still a problem in many scenarios of \ac{ML} solutions applied in cybersecurity, which may overlap with other areas, suggesting, in some cases, possible mitigation for them. As a study case, in most cases, we focus on malware detection or classification tasks \reviewdtrap{(for Android, Windows, and Linux operational systems)} given that they may contain all the problems listed. We want to acknowledge that we are not pointing fingers at anyone, given that our own work is subject to many problems stated here. \reviewdtrap{We also understand that, despite many problems stated, it does not mean that the findings of research papers are not useful from a practical perspective.} Our main contributions are the following: 

\reviewdtrap{
\begin{itemize}
    \item We show that many of the pitfalls are related to improperly considering the time when proposing solutions. This problem may be present in all the ML pipeline, from the data collection to the evaluation steps, and may be related to not considering the problem as a data stream.
    \item We advocate that more observational studies (i.e., research that focuses on analyzing ecosystem landscape, platforms, or specific types of attacks), are required so to the creation of useful datasets and solutions.
    \item We state that different approaches for attribute extraction impose varied costs, and might also lead to distinct ML outcomes. Thus, it is important to understand that each ML model and feature extraction algorithm serve different threat models, and it should be considered when evaluating a solution to avoid common pitfalls, such as applying the same criteria for online and offline applications.
    \item We show that concept drift detectors do no work in practice as intended in realistic experimental scenarios. Thus, new drift detection strategies that consider the challenges related to cybersecurity, such as the delay of labels and class imbalance, are needed to produce better solutions. 
    \item We point directions to future cybersecurity research works that make use of \ac{ML}, aiming to improve their quality to be used in real applications. To do so, we developed a novel checklist\footnote{The draft version of the checklist is available at 
    \blind.} to help the development of future ML solutions for cybersecurity based on the challenges, pitfalls, and problems reported in this work. Thus, anyone developing or reviewing a solution can use this checklist and get feedback reporting what could be improved or corrected according to our findings.
\end{itemize}}
This paper is organized as follows: first, in Section~\ref{sec:tasks}, we discuss how to identify the correct \ac{ML} task for a given cybersecurity problem; 
Further, this work is organized according to each step of the pipeline to develop ML solutions to cybersecurity (also presented in Section~\ref{sec:tasks}), including data collection (obtaining data for the ML solution), in Section~\ref{sec:data_collection}~, attribute extraction \review{(extracting metadata from the data previously obtained)} in Section~\ref{sec:attribute_extraction}, feature extraction \review{(extracting features from the attributes collected)} in Section~\ref{sec:feature_extraction}, model \review{(training and updating the model using the features extracted)} in Section~\ref{sec:model}, and evaluation \review{(evaluating the proposed solution)} in Section~\ref{sec:evaluation}. In Section~\ref{sec:discussion} we discuss how \ac{ML} applications should be understood, their limitations, and existing open gaps. Finally, we conclude our work in Section~\ref{sec:conclusion}.

\section{On the Modeling of Security Tasks as Machine Learning Problems}
\label{sec:tasks}

\ac{ML} has been applied in cybersecurity to solve different tasks. Table~\ref{tab:applications} presents illustrative examples we found in the literature on how different ML approaches can be used to solve different security problems. Our goal is not to present a new taxonomy or an exhaustive list of approaches, but to highlight that, on the one hand, (i) the same security problem might be addressed via different ML approaches and, on the other hand, (ii) the same ML solutions might be used for different security tasks and at different stages of a solution pipeline.

\begin{table}[!htpb]
\caption{\textbf{Representative Examples of Applications of \ac{ML} to Cybersecurity}. Distinct approaches are applied according to the specific security need. The same approach might be used to solve distinct security tasks. The placement of the solution in each pipeline step will present different pros and cons.}
\label{tab:applications}
\centering
    \begin{tabular}{ccc}
    \hline
    \textbf{Security Task} & \textbf{Specialized Task} & \textbf{\ac{ML} Task}               \\ \hline
    \multirow{3}{*}{\begin{tabular}[c]{@{}l@{}}Attack\\Detection\end{tabular}}             & Malware Detection   & Classification~\cite{arp2014}                 \\
                                             & Intrusion Detection & Outlier Detection~\cite{7092027}              \\
                                             & Spam Filtering      & Classification~\cite{6375166} \\ \hline
    \multirow{2}{*}{\begin{tabular}[c]{@{}l@{}}Incident \\Response\end{tabular}} & Malware Labelling   & Clustering~\cite{Sebastian2016}                     \\
                                             & Log Analysis        & Clustering~\cite{8622610} \\ \hline
    \multirow{2}{*}{\begin{tabular}[c]{@{}l@{}}Security\\Analysis\end{tabular}}  & Malware Analysis    & \begin{tabular}[c]{@{}c@{}}Reinforcement\\Learning/Ranking~\cite{ltr}\end{tabular} \\
                                             & Risk Assessment     & Regression~\cite{10.1145/3145905} \\ \hline
   \multirow{3}{*}{\begin{tabular}[c]{@{}l@{}}System\\Forensics\end{tabular}}              & Attribution         & Clustering~\cite{7555393} \\
                                             & Object Recognition  & \begin{tabular}[c]{@{}l@{}}Dimensionality\\Reduction~\cite{9067860}\end{tabular} \\ \hline 
    \end{tabular}
\end{table}

\noindent \textbf{Observation:} \ul{The same security problem presents different pros and cons depending on the adopted ML model.}
Attack detection tasks are typically modeled via binary classifiers that learn what are the benign and malicious classes. This strategy often achieves a reasonable detection level on the detection of known or similar attacks, but they are hardly ever able to detect new attacks (0 days). Modeling the attack detection problem as an outlier detection problem, in turn, tends to increase the ability to detect new threats. In this modeling, the ML model learns what is considered the normal behavior of the system and detects any deviation. A drawback of this approach in comparison to the typical classification problem is that it is hard to explain the detection result, as the model does not have knowledge of the attack class, but only about the normal profile. This difference might be key for choosing one or another approach for a given application, depending on the requirements for that scenario.

In some cases, the ML techniques present similar characteristics, such that they can be applied to multiple security tasks. For instance, clustering can be used as a classification approach in tasks ranging from (i) Adding malware variants to the same bucket for triaging; (ii) Adding security logs in a bucket of similar events; to (iii) Adding forensic evidence in similar buckets depending on the attacker authorship. In other cases, the nature of the ML solution must be as different from the others as the nature of the security task they need to solve. For instance, when malware analysts are reversing engineering samples, they want to know which strings are more informative about the malware nature. For that, they need a ranking algorithm, not a binary classifier.


\noindent \textbf{Challenge:} \ul{Selecting the proper ML task to solve the Security task.} Whereas multiple \ac{ML} tasks can be applied to security tasks, the identification of the right task for a given scenario is far from straightforward, and requires reasoning about multiple corner conditions, given that cybersecurity solutions are systems that may have multiple applications of ML integrated within a complex pipeline.
For instance, for the malware detection case, it is usual to find in the literature multiple solutions proposing a \ac{ML}-based engine to be applied by antiviruses (AVs). Most of these solutions are initially modeled as a classification problem (goodware vs. malware), which suffices for the detection task, but does not cover AV operation as a whole. In practice, AVs provide more than detection labels~\cite{BOTACIN2020101859}, they also attribute malware samples to families (e.g., ransomware, banker, and so on) to present a family label, which is essential to allow incident response procedures. Therefore, a \ac{ML} engine~\cite{BOTACIN2022102500} for an AV should also be modeled as a family attribution problem, which requires the application of both classifying and clustering algorithms~\cite{10.1109/TDSC.2017.2739145}.

\subsection{Machine Learning Pipeline for Cybersecurity}

\begin{figure}
    \centering
    \includegraphics[width=.75\columnwidth]{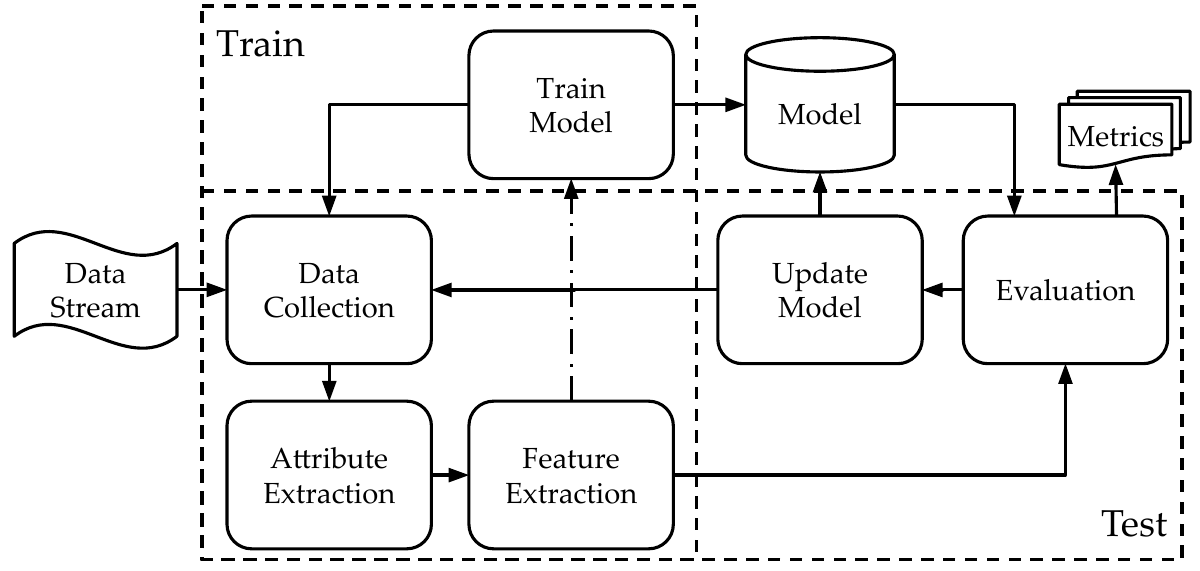}
    \caption{\textbf{Pipeline to develop and evaluate \acl{ML} solutions for cybersecurity.} \review{Each step is executed in sequence, considering that (i) the model is first trained and then (ii) tested/evaluated and updated (if needed).}}
    \label{fig:scheme}
\end{figure}

In Fig.~\ref{fig:scheme}, we show a pipeline to develop and evaluate \ac{ML} solutions \review{(both supervised and unsupervised methods)} for cybersecurity based on the literature, which is similar to any pattern classification task.
It consists of two phases: train, i.e., training a model with the data available at a given time \review{(for supervised methods, using the available labels and features to create decision boundaries, for unsupervised, using only the features to create clusters) and test, i.e., testing it considering the new data collected (for both supervised and unsupervised, using their labels -- to check if they were correctly predicted -- and features to update decision boundaries and clusters)}. \review{Note that we defined two steps after the data collection, given that, in many cybersecurity tasks, the raw data collected needs to have their metadata (attributes) extracted before actually being used by the ML model as features (attributes processed by a feature extractor). We understand that these steps may overlap in some cases, but for a more fine-grain discussion, we analyzed them separately.} \improve{Also, it is important to notice that security data, the input of the process, is available from a data stream that is in constant production, and the model and its metrics, the output of the process, are produced during the execution of the scheme.} \reviewdtrap{Thus, different from many works in the literature, we focus on problems related to data streams since they are close to real-world solutions for cybersecurity~\cite{androbin}.}

In the following sections, we discuss the challenges and pitfalls of multiple modeling strategies. Our goal is to help the reader to spot the quirks of ML applications at their multiple steps and abstraction layers.

\section{Data Collection Challenges}
\label{sec:data_collection}


The quality, quantity, and distribution of data inputted to a \acl{ML} algorithm (dataset) are the basis of an adequate learning process, since ML algorithms rely on the samples presented to them in the training step and the resulting model will allow further decision making. Thus, data collection, which comprises acquisition, enrichment/augmentation, and labeling~\cite{roh2019dc, Gron2017}, may be one of the most challenging steps of an ML project~\cite{8862913}. Besides real-world problems take these steps into consideration~\cite{gomes19ds}, certain research works might miss some of them. This might happen either due to the format of the dataset used or external reasons, such as privacy requirements about the data (might need differential privacy~\cite{dwork2014}), local laws that prevent the distribution of potentially harmful pieces of code, ease of accomplishing reproducibility of the experiments, and so on. Considering the steps performed in \ac{ML}-based cybersecurity systems~\cite{Saxe:2018:MDS:3299571} and to  provide a straightforward discussion, we assume that data can be available in three formats, which may overlap according to the task at hand or the chosen approach: 



\begin{itemize}
    \item \textbf{Raw Data:} data available in the same way they were collected, usually used to analyze their original behavior. For example: PE executables~\cite{nfs}, ELF~\cite{elfmalware}, APK packages~\cite{Allix:2016:ACM:2901739.2903508}, or network traffic data captured in PCAP format~\cite{pcapsource,pang2005first,hick2007caida,shiravi2012toward}. Compared to computer vision, the raw data are the images collected to create an image recognition model~\cite{hog2005}. In cybersecurity, AndroZoo~\cite{Allix:2016:ACM:2901739.2903508} provides many APK packages collected from several sources that could be used for malware detection systems or forensics analysis; 
    \item \textbf{Attributes:} filtered metadata extracted from the raw data with less noise and focus on the data that matters, being very suited for \ac{ML}. For example, CSV with metadata, execution logs of software or data extracted from its header~\cite{nfs, 2018arXiv180404637A}, or a summary of information from a subset of network traffic data. In computer vision, the attributes would be the gradient images extracted from the original ones~\cite{hog2005}. EMBER~\cite{2018arXiv180404637A} is a good example of a dataset containing attributes, available in JSON, extracted statically from raw data (Windows PE files). DREBIN~\cite{arp2014} is another example containing static attributes extracted from APK packages. 
    \item \textbf{Features:} features extracted from the attributes or raw data that distinguish samples, ready to be used in a classifier. For example the transformation of logs into feature vectors for every software mentioned in the previous item, whose positions of this vector correspond to the features~\cite{nfs}, or a transformation of traffic data into features containing the frequency of pre-determined attributes (i.e., protocols, amount of packages, bytes, etc.). In an image recognition problem, the features would be the histogram of gradients extracted using the gradient images created before~\cite{hog2005}. This type of data is usually used by researchers to make their experiments faster, given that they extract the features once and can share them to use as input for their models.
\end{itemize}

The data collection or even any of the dataset formats above are susceptible to problems that may affect the whole process of using \ac{ML} in any cybersecurity scenario and are frequently seen in the literature, such as data leakage (or temporal inconsistency), data labeling, class imbalance, and anchor bias.

\subsection{Data Leakage (Temporal Inconsistency)}
\label{subsec:data_leakage}


To evaluate a \ac{ML} system, it is common to split the dataset into at least two sets: one to train the model and another to test it. The \textit{k}-fold cross-validation is used to create \textit{k} partitions and evaluate a given model \textit{k} times, using one of the partitions as a test set and the remaining as a training set, taking the average of the metrics as a final result~\cite{Michie:1995:MLN:212782}. This is a common practice for many \ac{ML} experiments, such as image classification, where temporal information may not be important and is used in batches. However, when considering cybersecurity data, which are obtained from a stream, it is not real to consider data from different or mixed epochs to train and test a model (known as data leakage~\cite{kaufman2011}, data snooping~\cite{arpEtAl2022}, or temporal inconsistency), given that it could increase the detection accuracy because the model knows how future threats are (i.e., the model is exposed to test data when it was trained)~\cite{nfs}. For instance, consider a malware detector that works similarly to an antivirus, i.e., given a software, our model wants to know if it is malware or not to block an unwanted behavior. To create this model we train it using a set of malicious and benign software that are known and were seen in the past. Then, this model is used to detect if files seen in the future are malign, even if they perform totally different attacks than before. These new threats will just be used to update the model after they are known and labeled, which will probably increase the coverage of different unknown attacks, detecting more malware than before. 


Due to this temporal inconsistency problem, it is very important to collect the timestamp of the samples during the data collection and it is something not considered by some authors in cybersecurity. For instance, the DREBIN dataset~\cite{arp2014} (malware detection) makes all the malicious APKs available and also their attributes, but does not include the timestamp that they were found in the wild, making all the research works that use this dataset exposed to data leakage. We understand that sometimes it is difficult to set a specific release date for a sample, but they are needed to avoid this problem. For malware detection, for example, researchers usually use the date when 
they were first seen in VirusTotal as a timestamp~\cite{nfs, tesseract, Kantchelian2015, virustotal}.

\subsection{Data Labeling}
\label{subsec:data_labeling}

Labeling artifacts is essential to training and evaluating models. However, having artifacts properly labeled is as hard as collecting the malicious artifact themselves, as previously discussed. In many cases, researchers leverage datasets of crawled artifacts and make a strong hypothesis about them, such as ``all samples collected by crawling a blacklist are malicious``, or ``all samples collected from an App Store are benign``. These hypotheses are strong because previous work has demonstrated that even samples downloaded from known sources might be trojanized with malware~\cite{installers_dimva}. Therefore, a model trained based on this assumption would make the model also learn some malicious behaviors as legitimate. Also, some labels may be inaccurate, unstable, or erroneous, which may affect the overall classification performance of ML-based solutions~\cite{arpEtAl2022}.

A common practice to obtain labels and mitigate the aforementioned problems is to rely on the labels assigned by AVs by using the VirusTotal service~\cite{virustotal}, which provides detection results based on more than sixty antivirus engines available on the platform. Unfortunately, AV labels are not uniform~\cite{BOTACIN2020101859}, with each vendor reporting a distinct label. Therefore, researchers have either to select a specific AV to label their samples (according to their established criteria) or adopt a committee-based approach to unify the labels. For Windows malware, AVClass~\cite{Sebastian2016} is widely used for this purpose and, for Android malware, Euphony~\cite{Hurier2017} is used. Both of these techniques were evaluated by the authors using a high number of malware (8.9 million in AVClass and more than 400 thousand samples in Euphony), obtaining a significant F-measure score (bigger than 90\% in both cases) and generating consistent results to create real datasets such as AndroZoo~\cite{Allix:2016:ACM:2901739.2903508}, which uses Euphony~\cite{Hurier2017} to generate malware family labels. Although AVs can mitigate the labeling problem, their use should consider the drawbacks of AV internals. 
AVs provide two labels: detection and family attribution. Both the detection label as well as the family attribution label change very often over time: newly-released samples are often assigned a generic and/or heuristic label at the initial time and this is further evolved as new, more specific signatures are added to the AV to detect this threat class. Therefore, the date on which the samples are labeled might significantly affect the \ac{ML} results. Recent research shows that AVs labels might not establish before 20 or 30 days after a new threat is released~\cite{BOTACIN2020101859}. To mitigate this problem, delayed evaluation approaches should be considered, as shown in Section~\ref{sec:evaluation}.


\subsection{Class Imbalance}
\label{subsec:class_imbalance}

Class imbalance is a problem in which the data distribution between the classes of a dataset differs relatively by a substantial margin, and it is usually present in many research works~\cite{kaur2019systematic}. If we consider the Android landscape, the AndroZoo dataset contains only about 18\% of malicious apps (the remaining apps are considered benign~\cite{tesseract, Allix:2016:ACM:2901739.2903508}). This makes the Android malware detection problem more challenging due to the presence of imbalanced data. There are several methods in the literature whose aim is to overcome this problem by making use of  pre-processing techniques, improving the learning process (cost-sensitive learning), or using ensemble learning methods~\cite{kaur2019systematic, gomes19ds}. The two latter methods will be discussed in Section~\ref{sec:model}, since they are part of the process of training/updating the model. The former method (pre-processing) relies on resampling, i.e., removing instances from the majority classes (undersampling) or creating synthetic instances for the minority classes (oversampling)~\cite{gomes19ds}. 

\begin{figure*}
     \centering
     \begin{subfigure}[t]{0.32\textwidth}
         \centering
         \includegraphics[width=\textwidth]{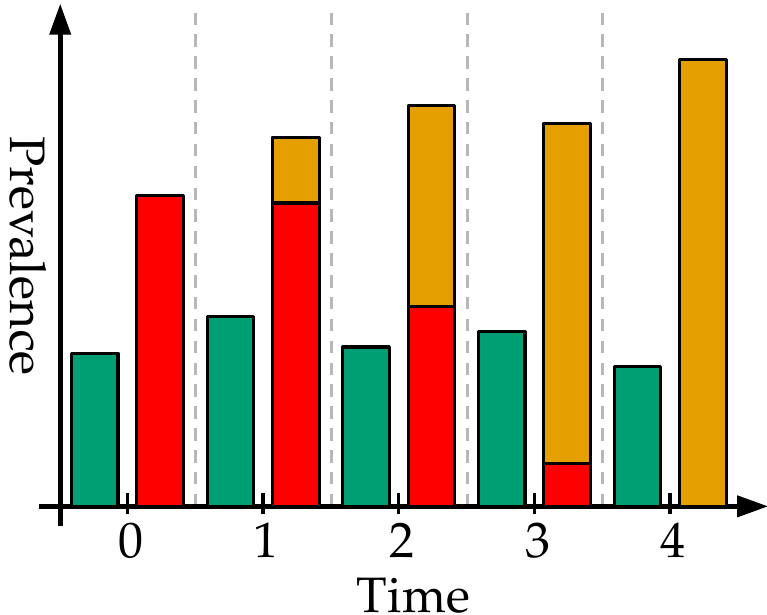}
         \caption{\textbf{No undersampling.} Dataset original distribution without undersampling.}
         \label{fig:undersampling_01}
     \end{subfigure}
     \hfill
     \begin{subfigure}[t]{0.32\textwidth}
         \centering
         \includegraphics[width=\textwidth]{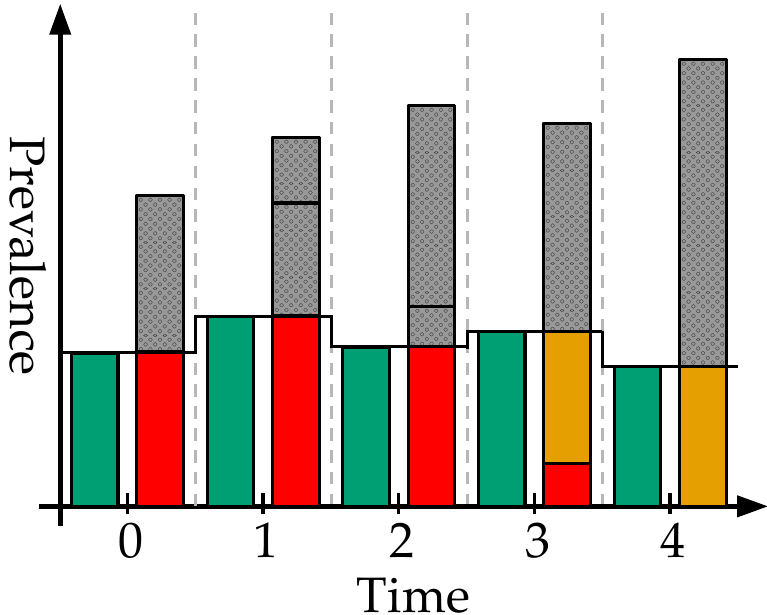}
         \caption{\textbf{Undersampling with no temporal information.} Undersampling dataset distribution when simply removing samples to keep the same distribution by class.}
         \label{fig:undersampling_02}
     \end{subfigure}
     \hfill
     \begin{subfigure}[t]{0.32\textwidth}
         \centering
         \includegraphics[width=\textwidth]{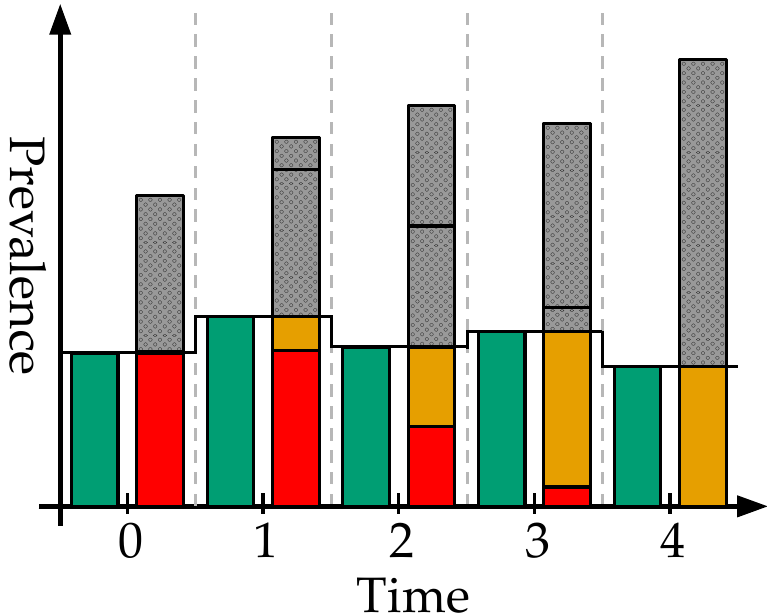}
         \caption{\textbf{Undersampling with temporal information.} Undersampling dataset distribution when keeping the same proportion of sub-classes of a given class.}
         \label{fig:undersampling_03}
     \end{subfigure}
        \caption{\textbf{Undersampling examples in a dataset.} Samples in green and red/orange represent different classes. Red and orange colors represent different sub-classes or concepts of the same class. The gray color with circles represents ignored/removed instances.}
        \label{fig:undersampling}
\end{figure*}

In the context of cybersecurity, undersampling may affect the dataset representation, given that removing some samples from a certain class can affect its detection. For instance, considering malware detection, removing malware samples may reduce the detection of some malware families (the ones that had samples removed) and also make their prevalence in a given time (monthly or weekly) less important than reality when creating a dataset, possibly not capturing a concept drift or sub-classes of the majority classes. Fig.~\ref{fig:undersampling} presents a \review{hypothetical dataset distribution} with two classes (green and red/orange), one of the classes has two concepts or two sub-classes (red and orange) as time goes by, i.e., consider it as being a dataset with normal and malicious behavior, the malicious one behavior that evolves or has sub-classes of behaviors as time goes by, \review{something that is known to happen in real-world cybersecurity datasets~\cite{nfs}}. In Fig.~\ref{fig:undersampling_01} we can see the original distribution of the dataset, with no technique applied. In Fig.~\ref{fig:undersampling_02} we can see the dataset distribution when removing samples (in gray) to keep the same distribution by class, which results in a different scenario than the reality since the orange concept or sub-class is not seen in some periods. In Fig.~\ref{fig:undersampling_03} we see the ideal scenario for undersampling when the proportion of concepts or sub-classes is the same for the same period while keeping the same number of samples for both classes (this may be a mitigation for the undersampling problem, but even with this solution important samples may be discarded).

\begin{figure*}
     \centering
     \begin{subfigure}[t]{0.32\textwidth}
         \centering
         \includegraphics[width=\textwidth]{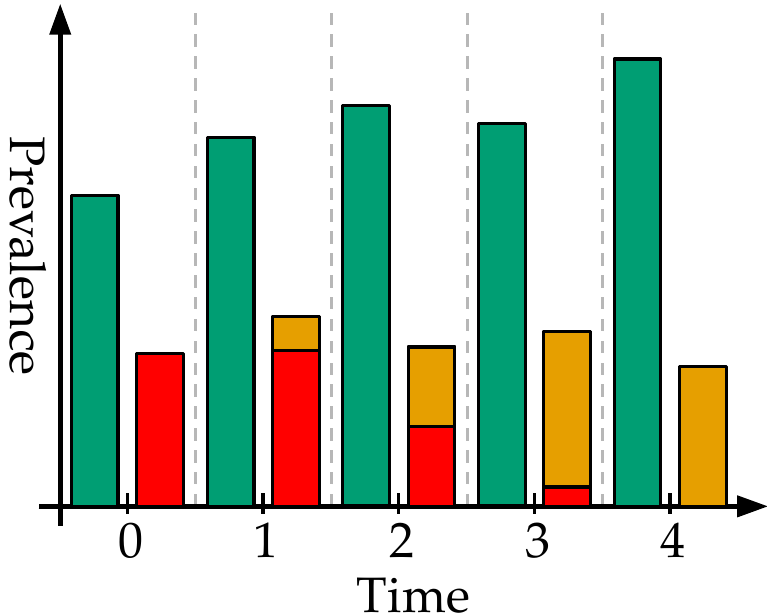}
         \caption{\textbf{No oversampling.} Dataset original distribution without oversampling.}
         \label{fig:oversampling_01}
     \end{subfigure}
     \hfill
     \begin{subfigure}[t]{0.32\textwidth}
         \centering
         \includegraphics[width=\textwidth]{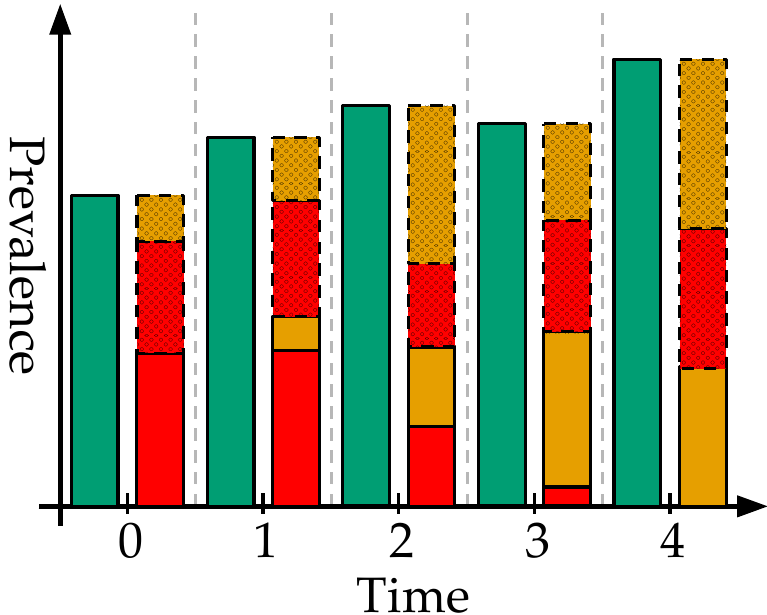}
         \caption{\textbf{Oversampling with no temporal information.} Oversampling dataset distribution without considering temporal information.}
         \label{fig:oversampling_02}
     \end{subfigure}
     \hfill
     \begin{subfigure}[t]{0.32\textwidth}
         \centering
         \includegraphics[width=\textwidth]{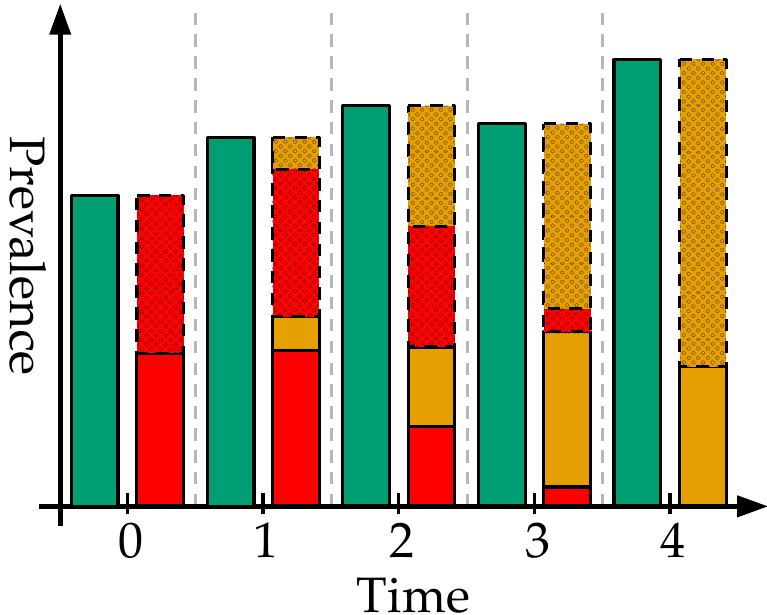}
         \caption{\textbf{Oversampling with temporal information.} Oversampling dataset distribution considering temporal information.}
         \label{fig:oversampling_03}
     \end{subfigure}
        \caption{\textbf{Oversampling examples in a dataset.} Samples in green and red/orange represent different classes. Red and orange colors represent different sub-classes or concepts of the same class. Samples with dashed line and circles are the synthetic instances created by oversampling.}
        \label{fig:oversampling}
\end{figure*}

An example of the oversampling technique is SMOTE, which consists in selecting samples that are close in the feature space, drawing a line between them, and generating new samples along with it~\cite{Chawla2002Smote}. Although such techniques are interesting, they may generate results that produce data leakage if they do not consider the time when creating synthetic samples. For instance, consider Fig.~\ref{fig:oversampling}, where we present again a \review{hypothetical dataset distribution} with two classes, with the same classes and problems as Fig.~\ref{fig:undersampling}. In the first case, in Fig.~\ref{fig:oversampling_01}, we see the original dataset distribution, where the class red/orange is the minority one. In Fig.~\ref{fig:oversampling_02} the problem of data leakage is shown: the artificial data generated (with dashed lines and circles) is based on all the dataset, without considering any temporal information. Thus, we can see the orange concept/sub-class at time 0 in the synthetic instances, which does not represent the real distribution of this class at that time (this problem happens all the time in this case, with the red concept/sub-class being shown even at time 4). In contrast, in Fig.~\ref{fig:oversampling_03} we can see an oversampling technique that considers the temporal information, generating synthetic data that correspond to the actual concept/sub-class of a given time, resulting in the same distribution of concepts/sub-classes as the original data. Despite also being an interesting approach, for the cybersecurity context, oversampling works at the feature level, which means that it may not generate synthetic raw data. For instance, if we consider these data as being applications, oversampling will only create synthetic feature vectors and not real applications that work.

\subsection{Dataset Size Definition}
\label{subsec:dataset_size_definition}

\begin{figure*}[!t]
  \centering
  \begin{minipage}[t]{0.49\textwidth}
    \includegraphics[width=\textwidth]{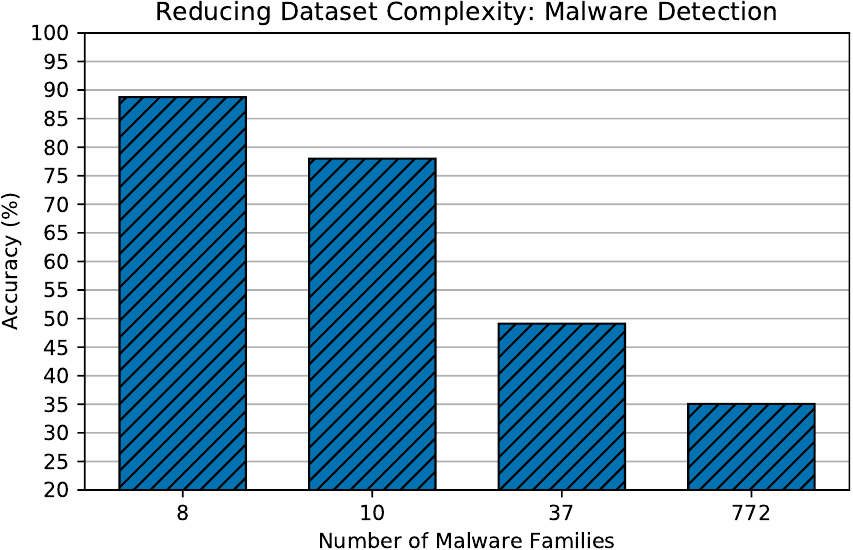}
    \caption{\textbf{Reducing dataset complexity.} The more the dataset is filtered, the bigger the accuracy achieved, which means that researchers must avoid filtered datasets to not produce misleading results~\cite{10.1007/978-3-030-30215-3_19}.}
    \label{fig:reducing_dataset_complexity}
  \end{minipage}
  \hfill
  \begin{minipage}[t]{0.49\textwidth}
    \includegraphics[width=\textwidth]{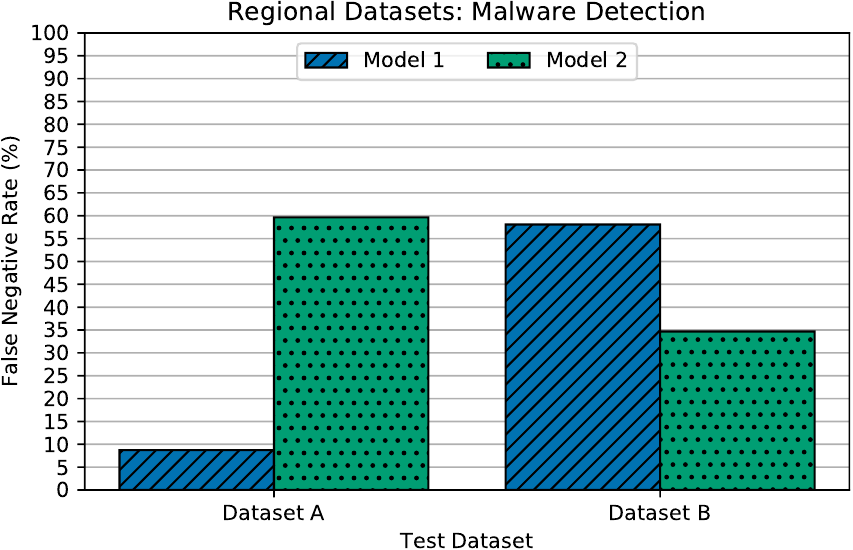}
    \caption{\textbf{Regional datasets.} Models may have a bias towards the scenario where the dataset used to train them was collected, indicating that they need to be specially crafted in some cases~\cite{10.1145/3433667.3433669}.}
    \label{fig:regional_datasets}
  \end{minipage}
\end{figure*}

A major challenge in creating a dataset is to define its size. Although it is a problem that affects all \ac{ML} domains, it has particularly severe implications in cybersecurity. On the one hand, a small dataset may not be representative of a real scenario, which invalidates the results of possible experiments using it. Also, some models may not generalize enough and not work as intended, presenting bad classification performance~\cite{Gron2017}. Limited evaluations are often seen in the cybersecurity context because collecting real malicious artifacts is a hard task, as most organizations do not share the threats that affect them to not reveal their vulnerabilities. 
On the other hand, a big dataset may result in long training time, and produce too complex models and decision boundaries (according to the classifier and parameters used) that are not feasible in reality (e.g., real-time models for resource-constrained devices), such as some deep learning models that usually requires a large amount of data to achieve good results~\cite{Najafabadi2015}. 
As an analogy, consider that a dataset is a map and, for instance, represents a city. It is almost impossible that this map has the same scale as the city, however, it can be useful by representing what is needed for a given purpose. For example, a map of public transportation routes is enough if our objective is to use it, but if we need to visit touristic places, this very same map will not be useful, i.e., new data or a new map is required~\cite{dataset_map}. These circumstances reflect the ideas of both George Box and Alfred Korzybski. Box said that "essentially, all models are wrong, but some are useful"~\cite{Shoesmith1987EmpiricalMA}, i.e., a model built using a dataset might be useful for a given task, but it will not be perfect -- there will be errors, we just need to know how wrong they have to be -- and it will not be good for other tasks, which makes necessary to collect more (or new) data and select new parameters for a new model. Korzybski mentioned that "the map is not the territory"~\cite{korzybski1931non}, meaning that it can be seen as a symbol, index, or representation of it, but it is not the place itself and it is prone to errors. In our case, a model or a dataset can represent a real scenario, but it is just a representation of it and may contain errors. These errors may present results that do not reflect reality. For instance, considering a malware detection task that uses grayscale images as representation for Windows binaries, when using a dataset that represents the real scenario with no data filtering, i.e., without removing specific malware families, the accuracy ($\approx$ 35\%) was much worse than other scenarios where the authors filter the number of malware families, reducing the complexity of the problem and achieving almost 90\% of accuracy, as shown in Fig.~\ref{fig:reducing_dataset_complexity}~\cite{10.1007/978-3-030-30215-3_19}.

Another point to consider when building a cybersecurity solution is that they may have regional characteristics that are reflected in the dataset and, as a consequence, in the model and its predictions~\cite{10.1145/3433667.3433669}. For instance, considering a malware detection task with two models that are based on classification trees: Model 1 (random forest with PE metadata~\cite{nfs}) is trained using data from region A (BRMalware dataset~\cite{nfs}) and Model 2 (LightGBM with PE metadata~\cite{2018arXiv180404637A}) is trained with data from region B (EMBER dataset~\cite{2018arXiv180404637A}). Ceschin et al. showed that when testing both models with test data from both regions (test dataset A, from region A, and test dataset B, from region B), they perform better in their respective regions and present a much higher false negative rate (FNR) in the opposite region (allowing malware to be executed if this solution is applied in a different region that it was designed for, for example), as shown in Fig.~\ref{fig:regional_datasets}. Thus, it is important to consider collecting new data when implanting a known solution to a new target scenario (or region).

\begin{figure*}[!t]
     \centering
     \begin{subfigure}[t]{0.49\textwidth}
         \centering
         \includegraphics[width=\textwidth]{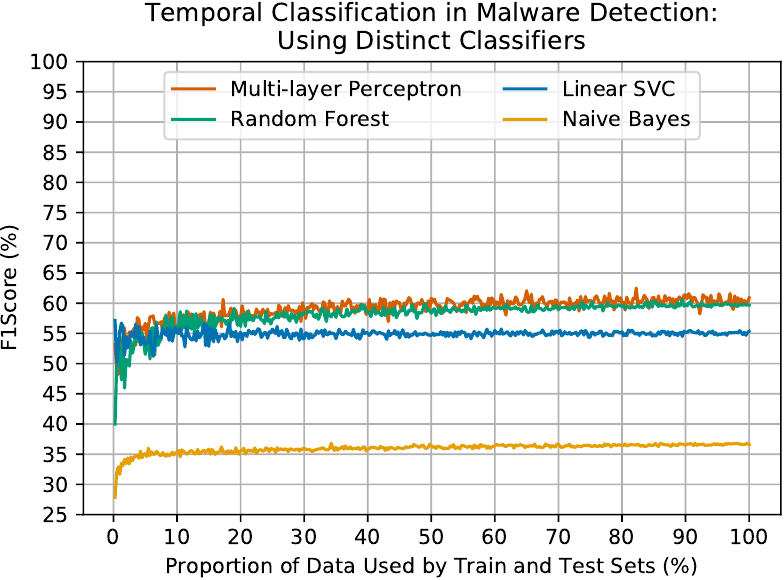}
         \caption{\textbf{Comparing classifiers.} ``Optimal'' classification performance is achieved by using only around 10\% to 20\% of the original dataset in all cases.}
         \label{fig:dataset_size_classifiers}
     \end{subfigure}
     \hfill
     \begin{subfigure}[t]{0.49\textwidth}
         \centering
         \includegraphics[width=\textwidth]{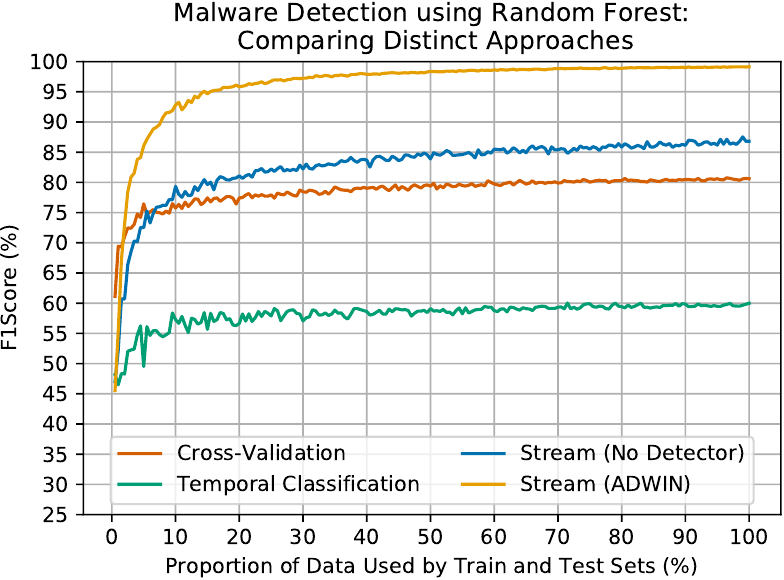}
         \caption{\textbf{Comparing approaches.} All of them stabilize their f1score after 30\% to 40\% proportions, less than half of the original dataset.}
         \label{fig:dataset_size_approaches}
     \end{subfigure}

    \caption{\textbf{Dataset size definition in terms of f1score.} In both experiments a similar behaviour is seen, with a certain stability in classification performance after using only a given proportion of the original dataset.}
    \label{fig:dataset_size}
\end{figure*}

To elaborate our discussion about dataset size definition, we created two experiments to better understand how much data we need to achieve representative results using a subset of AndroZoo dataset~\cite{androbin, Allix:2016:ACM:2901739.2903508} for malware detection, composed of $347,444$ samples ($267,342$ benign and $80,102$ malicious applications) spanning from 2016 to 2018. Both experiments consist in understanding how much data we need to achieve a stable classification performance based on the dataset proportion used to train our models. To do so, we divided the dataset by months and reduced the proportion of goodware and malware in both the training and test set (the first half of the dataset, ordered by their first seen date in virus total, is used to train and the second, to test). Thus, in the first experiment, we tested different classifiers using temporal classification (using the training set with ``known data'' to train the models and ''future data`` to test them), with different proportions of data in the training and test dataset, to see how each one of them was going to perform. Surprisingly, all classifiers \review{(Multi-layer Perceptron, Linear SVC, Random Forest, and Naive Bayes~\cite{scikit-learn})} had similar behaviors, also presenting similar curves as consequence and reaching an ``optimal'' classification performance by using only around 10\% to 20\% of the original dataset, with multi-layer perceptron and random forest achieving the best overall results, as shown in Fig.~\ref{fig:dataset_size_classifiers}. It is clear to see that, after these proportions, the f1score was almost stable, improving just a little even when increasing a lot the amount of data used. Furthermore, in the second experiment, we compared different approaches in a unique classifier \review{(Random Forest)} to see if they present the same behavior as the first one. To do so, we used only random forest in four different approaches: cross-validation (randomly selecting train and test samples, with data leakage), temporal classification (the same approach as the first experiment), and stream with and without drift detector (initializing a stream version of the random forest~\cite{gomes2017adaptive} -- known as adaptive random forest -- with the training data and incrementally updating it with testing data, using the ADWIN drift detector~\cite{adwin} to detect changes in data distribution when it is enabled). As we can see in Fig.~\ref{fig:dataset_size_approaches}, all the approaches also presented similar behavior and curves, but this time, all of them started to stabilize their f1score after 30\% to 40\% proportions, which means that they just need at least half of the dataset to present almost the same result as using the entire dataset. With both experiments, we conclude that, at some point, it may be more important to look for new features or representation strategies than to add more data to the training set, given that the classification performance is not improved so much according to our experiments.


The implication of these findings for cybersecurity is that more observational studies -- research that focuses on analyzing ecosystem landscape, platforms, or specific types of attacks to inform the development of future solutions -- are required so to allow for the creation of useful datasets and \ac{ML} solutions. Due to the parsimonious number of this kind of study, one might fall for the Anchor bias~\cite{doi:10.1111/j.1467-9280.2006.01704.x}, a cognitive bias where an individual relies too heavily on an initial piece of information offered (the ``anchor'') during decision-making to make subsequent judgments. When the value of the anchor is set, future decisions are made using it as a baseline. For instance, if we consider research that uses a one million samples dataset, it is going to become the anchor for future research, even if this dataset is not consistent with the real scenario. Thus, choosing or creating a dataset is not about its size, but its representation of the real world for the task being performed, as a map. In addition, concept drift and evolution are the nature of cybersecurity datasets, which makes it necessary to collect samples from different epochs to correctly evaluate any solution~\cite{nfs,GIBERT2020102526}. We acknowledge here that some approaches need more data to achieve better results, such as deep learning techniques~\cite{239524}, and it may be a limitation for them. A good example of building a real dataset is the one built by Pendlebury et al. for malware detection, where the proportion of samples found in the dataset is the same found in the wild by AndroZoo collection of Android applications~\cite{tesseract,Allix:2016:ACM:2901739.2903508}. 






\section{Attribute Extraction}
\label{sec:attribute_extraction}

Extracting attributes, i.e., selecting filtered metadata collected from the raw data, is a key step to creating useful features for the \ac{ML} models. In this section, we  pinpoint 
 the impact of different attributes in \ac{ML} solutions. \review{Note that we have separate definitions for attributes and features (see Section~\ref{sec:data_collection}): the former is filtered metadata extracted from the raw data obtained in the data collection step and \improve{strictly related to the security task being performed}, whereas the latter is the attributes transformation into a distinct set of samples representation ready to serve as input to a model. They usually cannot be directly used by an ML model, given that they need to be preprocessed and/or transformed into features (generally, numerical) to be used as input of a classifier.}

\subsection{The Impact of Different Attributes} 
\label{subsec:impact_attributes}



\begin{figure}[ht]
     \centering
     \includegraphics[width=.5\textwidth]{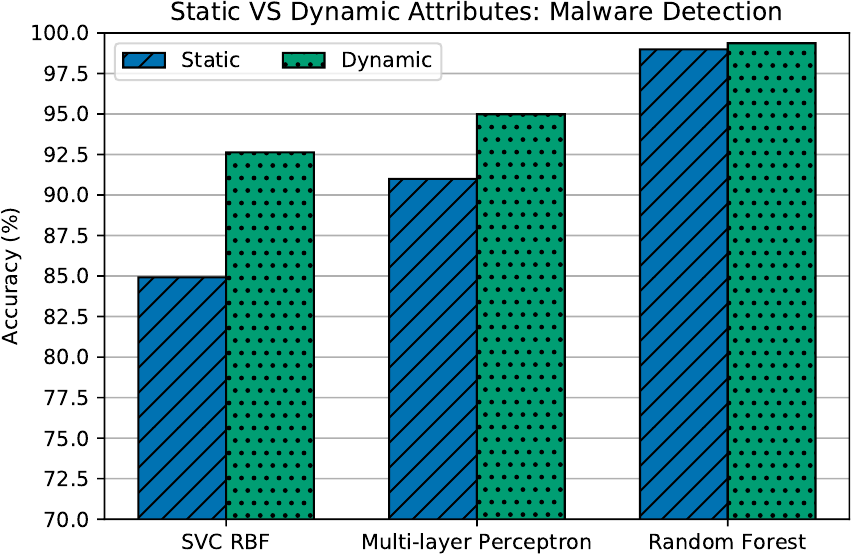}
     \caption{\textbf{Static VS Dynamic attributes in malware detection.} According to the classifier used, accuracy is highly impacted by the type of attributes used (experiments derived from Galante et al.~\cite{beyondaccuracy}).}
     \label{fig:static_dynamic_attributes}
\end{figure}

Different approaches for attribute extraction impose varied costs, and might also lead to distinct \ac{ML} outcomes. Naturally, executing an artifact to extract feature costs more than statically inspecting it, but the precision of the classification approach might get higher if this data is properly used. Therefore, the selection of the attribute extraction procedure should consider its effect on the outcome. 
Galante et al. show the effect of attribute extraction procedures over three distinct pairs of models to detect Linux malware (SVM with RBF kernel, Multi-layer Perceptron, and Random Forest)~\cite{beyondaccuracy}. These three pairs of models consider the same set of features, \improve{one of them considering statical attributes and the other, dynamic attributes.}
For instance, in statically extracted attributes, the authors consider the presence of the \texttt{fork} system call in the import table of the sample's binary to build a feature vector, whereas, in the corresponding dynamic attributes, they consider the frequency of the fork system call invocation during the sample execution in a sandbox to build a feature vector as well. 
%
The authors
used a balanced dataset of Linux binaries with benign and malign applications as input to all of the aforementioned models and the outcome was that the dynamic extraction approach outperformed the static approach, as shown in the accuracy rates (Fig.~\ref{fig:static_dynamic_attributes}). Although the dynamic attributes greatly impacted SVM and Multi-layer Perceptron results, it did not hold for Random Forest -- the difference between feature vectors resulting from dynamic and static attributes was not significant, which means that using only static attributes is enough for this particular model and scenario. 


Nguyen et al. compared four different attribute extraction methods for malware detection~\cite{Nguyen2022}: raw bytes~\cite{roots_model1}, EMBER features (static PE file header attributes)~\cite{2018arXiv180404637A}, CAPA features~\cite{capa}, and dynamic analysis. While the raw bytes attributes take $0.002$ seconds per file to be classified, the static attributes (EMBER) take $0.09$ seconds, the CAPA features takes $45.75$ seconds, and the dynamic analysis takes $526$ seconds per file, i.e., using raw bytes on a file is over $26,300$ times faster than running dynamic analysis. Finally, according to their experiments, the raw bytes model achieved a higher malware detection accuracy than the dynamic analysis ($\approx 90\%$ vs $\approx 85\%$), while the ensemble using both of them achieves almost $93\%$, showing that, by combining different features, it is possible to improve classification performance considering that it would be much more expensive. 




\section{Feature Extraction Pitfalls}
\label{sec:feature_extraction}

It is faster and simpler to use numerical or categorical attributes in any model, either by just encoding the categorical ones, or normalizing both of them. However, these attributes may not be directly used in the \ac{ML} model, depending on their type after being extracted from raw data. For instance, a list of system calls or libraries used by software must pass through one more processing step before it can serve as input to the ML algorithm. This step is known as feature extraction, and its goal is to transform these attributes into something readable by the classifier and simplify the data while keeping the level of provided information, but reducing the number of resources used to describe them~\cite{Gron2017, Saxe:2018:MDS:3299571}. Thus, there are several approaches to extracting features from attributes, and they usually rely on well-known techniques from ML literature, such as text classification, image recognition or classification, graph feature learning, and deep learning~\cite{GIBERT2020102526, ijcai2020-668}. \improve{Finally, according to the feature extraction selected, an ML solution may present several challenges and pitfalls.}

\subsection{Adapting to Changes}
\label{subsec:adapting_changes_features}

Many of the feature extractors mentioned in the literature
need to be created based on a training dataset to, for instance, create a vocabulary (e.g., TF-IDF~\cite{Jones72astatistical}) or to compute the weights of the neural network used (e.g., Word2Vec~\cite{google_word2vec1}), similar to an \ac{ML} model. Thus, as time goes by, it is necessary to update the feature extractor used if a concept drift or evolution happens in the application domain, which is something very common in cybersecurity environments due to new emerging threats~\cite{10.1145/2517312.2517320, nfs, GIBERT2020102526}. For instance, when using any vocabulary-based feature extractor for malware detection based on static features, such as the list of libraries used by a software, new libraries may be developed as time goes by (and they are not present in the vocabulary), which can result in a concept drift and make the representation created for all the new software outdated. In response to concept drift, the feature extractor may also need to be updated when it is detected and not only the classifier itself, requiring an efficient performance to update both of them. 


\begin{figure*}[!ht]
     \centering
     \begin{subfigure}[t]{0.49\textwidth}
         \centering
         \includegraphics[width=\textwidth]{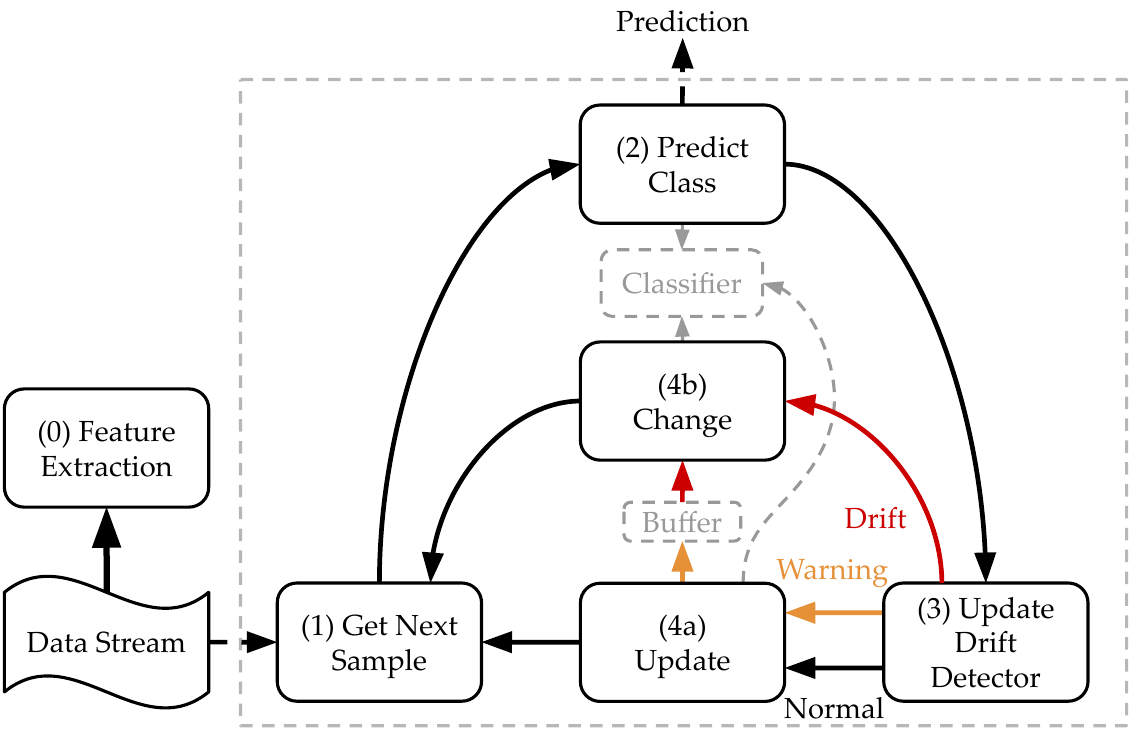}
         \caption{\textbf{Traditional data stream pipeline.} Feature extraction is applied to the raw data stream data before using them in the learning cycle. Then, the classifier generates the prediction of the current sample and uses it to update the model, adding it to a buffer or changing the model according to the concept drift detector level.}
         \label{fig:datastream_old}
     \end{subfigure}
     \hfill
     \begin{subfigure}[t]{0.49\textwidth}
         \centering
         \includegraphics[width=\textwidth]{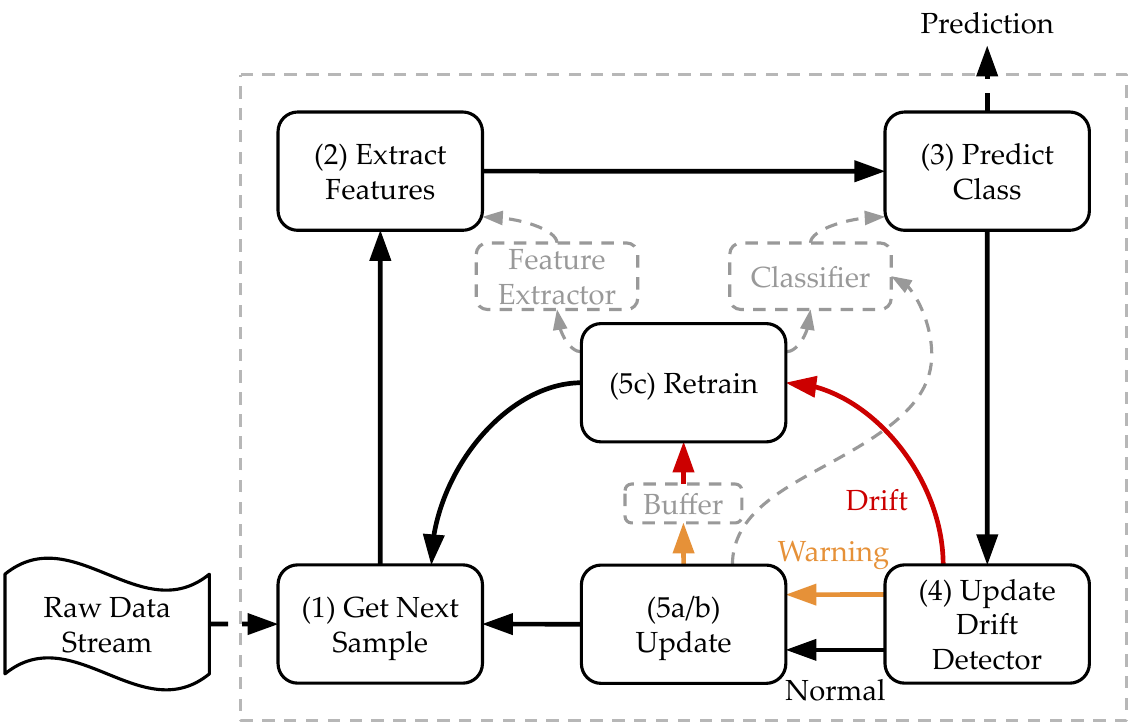}
         \caption{\textbf{New data stream pipeline with feature extractor.} Every time a new sample is obtained from the raw data stream, its features are extracted and presented to a classifier, generating a prediction, which is used by a drift detector that defines the next step: update the classifier or retrain both the classifier and feature extractor~\cite{androbin}.}
         \label{fig:datastream_new}
     \end{subfigure}

    \caption{\textbf{Data stream pipeline.} Comparing the traditional data stream pipeline with possible mitigation that adds the feature extractor in the process, generating updated features as time goes by~\cite{androbin}.}
    \label{fig:datastream}
\end{figure*}


To illustrate this challenge, Ceschin et al. created an experimental scenario using a proportionally reduced version of the AndroZoo dataset~\cite{Allix:2016:ACM:2901739.2903508} with almost 350K samples (the same we used in Section~\ref{subsec:dataset_size_definition}), and the DREBIN dataset~\cite{arp2014}, composed of $129,013$ samples ($123,453$ benign and $5,560$ malicious Android applications)~\cite{androbin}. The authors sorted the samples by their first seen date in VirusTotal~\cite{virustotal} and trained two base Adaptive Random Forest classifiers~\cite{gomes2017adaptive} with the first year of data, both of them include the ADWIN drift detector~\cite{adwin}, which usually has the best classification performance in the literature. When a concept drift is detected, the first classifier is updated using always the same features from the start, while the second one is entirely retrained from scratch, updating not only the classifier but also the feature extractor. In a traditional data stream learning problem that includes concept drift, the classifier is updated with new samples, which already had their features extracted previously using a feature extractor, when a change occurs (generally the ones that created the drift), as shown by 
steps in Fig.~\ref{fig:datastream_old}~\cite{Gama:2014:SCD:2597757.2523813}.
Alternatively, the data stream pipeline proposed in this experiment as mitigation to this problem also considers the feature extractor under changes, retraining both feature extractor and classifier according to the following five steps in Fig.~\ref{fig:datastream_new}~\cite{androbin}.

\begin{figure*}[!ht]
     \centering
     \begin{subfigure}[t]{0.49\textwidth}
         \centering
         \includegraphics[width=\textwidth]{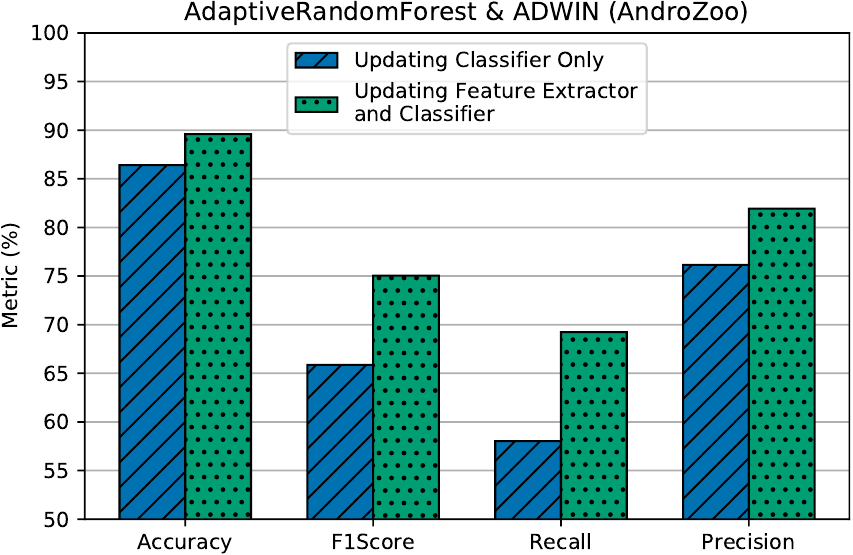}
         \caption{\textbf{AndroZoo dataset.} Adaptive Random Forest with ADWIN when classifying AndroZoo~\cite{androbin}.}
         \label{fig:adapting_features_androzoo}
     \end{subfigure}
     \hfill
     \begin{subfigure}[t]{0.49\textwidth}
         \centering
         \includegraphics[width=\textwidth]{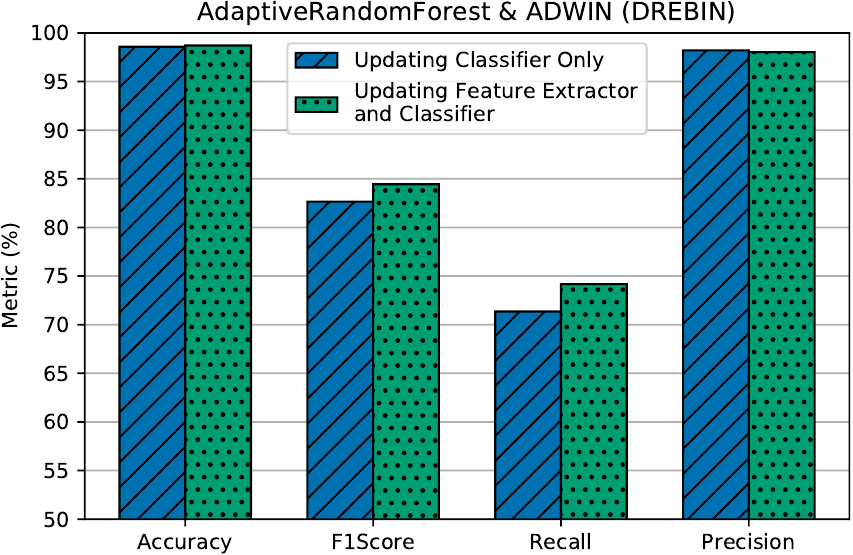}
         \caption{\textbf{DREBIN Dataset.} Adaptive Random Forest with ADWIN when classifying DREBIN~\cite{androbin}.}
         \label{fig:adapting_features_drebin}
     \end{subfigure}

    \caption{\textbf{Adapting features improves classification performance.} When considering the feature extractor in the pipeline, updating it when drift occurs is better than using a static representation (using a unique feature extractor based on the first training set). Experiments derived from Ceschin et al.~\cite{androbin}.}
    \label{fig:adapting_features}
\end{figure*}

Fig.~\ref{fig:adapting_features} shows the results presented by the authors regarding the impact on classification performance caused when the feature extractor is updated after a concept drift is detected (AndroZoo dataset in Fig.~\ref{fig:adapting_features_androzoo} and DREBIN dataset in Fig.~\ref{fig:adapting_features_drebin}), improving by almost 10 percentage points of f1score when classifying AndroZoo dataset (from 65.86\% to 75.05\%). Thus, we emphasize the importance of including the feature extractor in the incremental learning process~\cite{androbin}.


\subsection{Adversarial Features (Robustness)}
\label{subsec:adversarial_ml_features}


There are multiple ways to choose features that represent the samples involved in a security problem, and the final accuracy is not the only metric that should be considered during the choosing step. Another important point to take into account is the robustness of the resulting model against adversarial machine learning: attackers may try to adapt their malicious samples to make their features look similar to benign samples while maintaining the same original behavior~\cite{10.1145/3375894.3375898}. Thus, researchers must  think like an attacker when choosing which attributes and features will be used to build an \ac{ML} system for cybersecurity, given that some of them might be easily changed to trick the \ac{ML} model.

\begin{figure*}[!htpb]
    \centering
    \includegraphics[width=\linewidth]{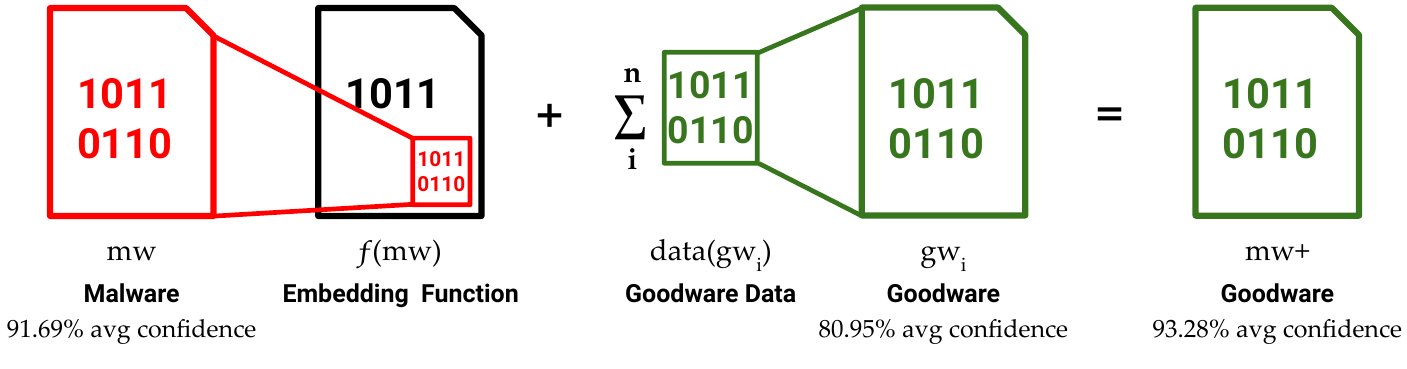}
    \caption{\textbf{Adversarial malware generation.} It is possible to change classifiers' output by just using an embedding function to add
    malware payloads within a new file and adding goodware data to it, such as strings and bytes 
    from a set of goodware~\cite{10.1145/3375894.3375898}.}
    \label{fig:adversarial_malware}
\end{figure*}

To illustrate the impact of adversarial machine learning in cybersecurity solutions, let's consider the following malware detection models proposed in the academic literature: (i) \texttt{MalConv}~\cite{roots_model1} and \texttt{Non-Negative MalConv}~\cite{roots_model2}, which are deep learning classifiers whose features are the raw bytes of an input file; and (ii) \texttt{LightGBM}~\cite{roots_lightgbm}, whose features consist of a matrix created using a hashing trick and histograms based on the inputted binary files characteristics (PE header information, file size, timestamp, imported libraries, strings, etc.). Both of these models can be easily bypassed using simple strategies that create totally functional adversarial malware. The formers (raw bytes-based models) can be tricked by simply appending goodware bytes or strings present in goodware applications at the end of the malware binary. The performed appendage does not affect the original binary execution and biases the detector toward the statistical identification of these goodware features. The latter (file characteristics-based model) can be bypassed by embedding the original binary into a generic file dropper, which extracts the embedded malicious content in runtime and executes it. The dropping technique, presented in Fig.~\ref{fig:adversarial_malware}, bypasses detection because the classifier would inspect the external dropper (only contains characteristics of benign applications, such as headers) instead of the embedded payload (the de facto malicious application). Previous work showed that the combination of the aforementioned counter-ML strategies can generate adversarial malware capable of bypassing both types of detection models, as well as affect the detection rate of antiviruses that rely on \ac{ML} in their engines (as shown in Fig.~\ref{fig:adversarial_malware}, where a malware with 91.69\% average confidence is transformed into a goodware with 93.28\% of average confidence)~\cite{10.1145/3375894.3375898}.

Other types of attributes may be more resistant to these attacks. Although approaches based on Control Flow Graphs (CFGs) are not affected by the appending of extra, spurious bytes, they can be bypassed by malware samples that modify their internal control flow structures to resemble legitimate software pieces~\cite{CALLEJA2018113}. A strategy to handle this problem is to select the features that are more resistant to modifications on the original binary (e.g., the use of loop instructions whose code is enough to distinguish malicious programs, followed by the building of a feature space containing a set of labels for each of them, thus making adversarial feature vectors more difficult to attackers~\cite{10.1145/3274694.3274731}).


In summary, we advocate for more studies about the robustness of features for cybersecurity, given that it is something crucial for the development of real applications. Thus, the aphorism ``Garbage In, Garbage Out'' used in many ML contexts is also valid for the quality of a solution, since it may become useless if subject to successful adversarial attacks. Proper \ac{ML} models require high-quality training data and robust features in order to produce high-quality and robust classifiers~\cite{10.1145/3351095.3372862}. Security-wise, it is important to understand that each \ac{ML} model and feature extraction algorithm serve different threat models. Therefore, the resistance of a feature to a given type of attack should be evaluated considering the occurrence, prevalence, and impact of this type of attack in the specific application scenario (e.g., newly-released binaries being distributed with no validation codes, such as signatures and/or MAcs are more prone to be vulnerable to random data append attacks than the cases in which the original binary integrity is verified).





\section{ML Modelling Issues and Solutions}
\label{sec:model}


An \ac{ML} model is a mathematical model that generates predictions by finding relationships between patterns of the input features and labels in the data~\cite{model_aws20}. Thus, when using machine learning for any task, it is common to test different types of models and fine-tune them to find the one that best suits the application~\cite{bishop:2006}. In cybersecurity, due to the dynamic scenarios presented in many tasks, streaming data models are strongly recommended to achieve a good performance, given that they belong to non-stationary distributions, new data are produced all the time, and they can be easily updated or adapted with them~\cite{MOA-Book-2018}. As a consequence, it is important to understand how to effectively use and, sometimes, implement an \ac{ML} model in these scenarios, given that they may present many drawbacks that are not feasible in a real application.

\subsection{Concept Drift and Evolution}
\label{subsec:concept_drift_evolution}

\begin{figure*}[!ht]
     \centering
     \begin{subfigure}[t]{0.49\textwidth}
         \centering
         \includegraphics[width=\textwidth]{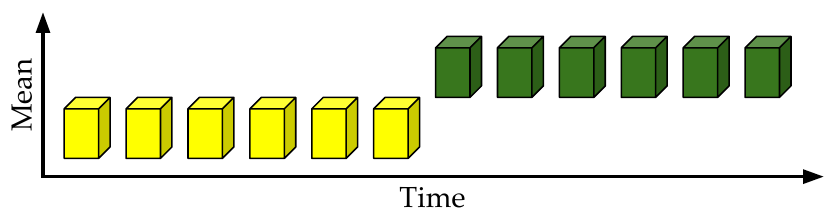}
         \caption{\textbf{Sudden drift.} A concept is suddenly replaced by a new one.}
         \label{fig:sudden_drift}
     \end{subfigure}
     \hfill
     \begin{subfigure}[t]{0.49\textwidth}
         \centering
         \includegraphics[width=\textwidth]{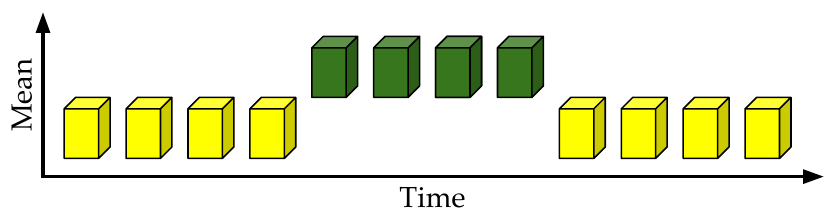}
         \caption{\textbf{Recurring concepts.} Previous active concept reappears after some time.}
         \label{fig:recurring_concepts}
     \end{subfigure}
     
     \begin{subfigure}[t]{0.49\textwidth}
         \centering
         \includegraphics[width=\textwidth]{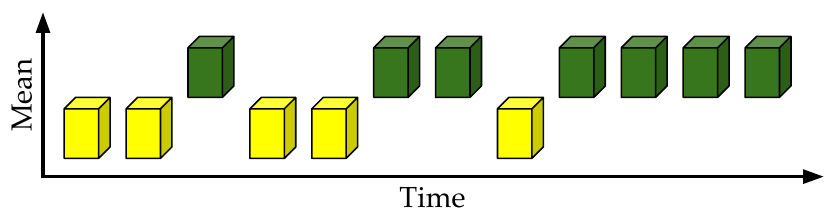}
         \caption{\textbf{Gradual drift.} The probability of finding the previous concept decreases and the new one increases until it is completely replaced.}
         \label{fig:gradual_drift}
     \end{subfigure}
     \hfill
     \begin{subfigure}[t]{0.49\textwidth}
         \centering
         \includegraphics[width=\textwidth]{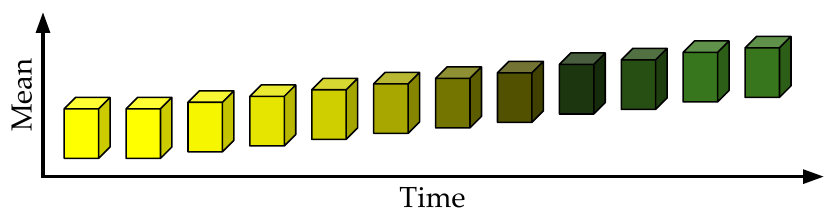}
         \caption{\textbf{Incremental drift.} The difference between the old concept and the new one is very small and the difference is only noticed when looking at a longer period.}
         \label{fig:incremental_drift}
     \end{subfigure}

    \caption{\textbf{Drift types.} Different types of concept drift presented in the literature~\cite{Lemaire2015}.}
    \label{fig:drift_types}
\end{figure*}

Concept drift is the situation in which the relation between the input data and the target variable (the variable that needs to be learned, such as class or regression variable) changes over time~\cite{Gama:2014:SCD:2597757.2523813}. It usually happens when there are changes in a hidden context, which makes it challenging since this problem spans different research fields~\cite{WANG2011247}. In cybersecurity, these changes are caused by the arms race between attackers and defenders, once attackers are constantly changing their attack vectors when trying to bypass defenders' solutions~\cite{10.1145/3375894.3375898}. In addition, concept evolution is another problem related to this challenge, which refers to the process of defining and refining concepts, resulting in new labels according to the underlying concepts~\cite{kulesza2014_concept_evolution}. Thus, both problems (drift and evolution) might be correlated in cybersecurity, given that new concepts may result in new labels, such as new types of attacks produced by attackers.
As shown in Fig.~\ref{fig:drift_types}, there are four types of concept drift according to the literature: (i) sudden drift, when a concept is suddenly replaced by a new one; (ii) recurring concepts, when a previous active concept reappears after some time; (iii) gradual drift, when the probability of finding the previous concept decreases and the new one increases until it is completely replaced; and (iv) incremental drift, when the difference between the old concept and the new one is very small and the difference is only noticed when looking at a longer period~\cite{Lemaire2015}. In security contexts, a sudden drift is when an attacker creates a totally new attack; gradual drift is when new types of attacks are created and replace previous ones; the recurring concept is when an old type of attack starts to appear again after a given time; and incremental drift is when the attackers make few modifications in their attacks in a way that their concepts change over a large period.

Despite being considered a challenge in cybersecurity~\cite{GIBERT2020102526}, few works addressed both problems in the literature. For instance, Masud et al., to the best of our knowledge, were the first to treat malware detection as a data stream classification problem and mention concept drift. The authors proposed an ensemble of classifiers that are trained from consecutive chunks of data using $v$-fold partitioning of the data, reducing classification error compared to other ensembles and making it more resistant to changes when classifying real botnet traffic data and real malicious executables~\cite{Masud2008}.
Singh et al. proposed two measures to track concept drift in static features of malware families: relative temporal similarity and meta-features~\cite{Singh2012}. The former is based on the similarity score (cosine similarity or Jaccard index) between two time-ordered pairs of samples and can be used to infer the direction of the drift. The latter summarizes information from a large number of features, which is an easier task than monitoring each feature individually. 
Narayanan et al. presented an online \ac{ML}-based framework named DroidOL to handle it and detect malware~\cite{Narayanan2016}. To do so, they use inter-procedural control-flow sub-graph features in an online passive-aggressive classifier, which adapts to the  malware drift and evolution by updating the model more aggressively when the error is large and less aggressively when it is small. They also propose a variable feature-set regimen that includes new features to samples, including their values when present and ignoring them when absent (i.e., their values are zero).
Deo et al. proposed the use of Venn-Abers predictors to measure the quality of binary classification tasks and identify antiquated models, which resulted in a framework capable of identifying when they tend to become obsolete~\cite{Deo2016}. 
Jordaney et al. presented Transcend, a framework to identify concept drift in classification models which compares the samples used to train the models with those seen during deployment~\cite{Jordaney2017}. To do it, their framework uses a conformal evaluator to compute algorithm credibility and confidence, capturing the quality of the produced results that may help to detect concept drift. 
Anderson et al. showed that, by using reinforcement learning to generate adversarial samples, it is possible to retrain a model and make these attacks less effective, also protecting it against possible concept drift, given that it hardens a machine learning model against worst-case inputs~\cite{anderson2018learning}. 
Xu et al. proposed DroidEvolver, an Android malware detection system that can be automatically updated without any human involvement, requiring neither retraining nor true labels to update itself~\cite{Xu2019}. The authors use online learning techniques with evolving feature sets and pseudo labels, keeping a pool of different detection models and calculating a juvenilization indicator, which determines when to update its feature set and each detection model. 
Finally, Ceschin et al. compared a set of Windows malware detection classifiers that use batch machine-learning models with ones that take into account the change of concept using data streams, emphasizing the need to update the decision model immediately after a concept drift is detected by a concept drift detector, which are state-of-the-art techniques used in the data stream learning literature~\cite{nfs}. The authors also show that the malware concept drift is strictly related to their concept evolution, i.e., due to the appearance of new malware families.



In contrast, data stream learning literature already proposed some approaches to deal with concept drift and evolution, called concept drift detectors, that, to the best of our knowledge, were not totally explored by cybersecurity researchers. There are supervised drift detectors that take into account the ground-truth label to make a decision and unsupervised ones that do not. DDM (Drift Detection Method~\cite{gama2004}), EDDM (Early Drift Detection Method~\cite{baena2006}) and ADWIN (ADaptive WINdowing~\cite{adwin}) are examples of supervised approaches. Both DDM and EDDM are online supervised methods based on sequential error (prequential) monitoring, where each incoming example is processed separately estimating the prequential error rate. This way, they assume that the increase in consecutive error rate suggests the occurrence of concept drifts. DDM directly uses the error rate, while EDDM uses the distance error rate, which measures the number of examples between two classification  errors~\cite{baena2006}. These errors trigger two levels: warning and drift. The warning level suggests that the concept starts to drift, updating an alternative classifier using the examples which rely on this level. The drift level suggests that the concept drift occurred, and the alternative classifier built during the warning level replaces the current classifier. ADWIN keeps statistics from sliding windows of variable  size, which are used to compute the average of the change observed by cutting these windows at different points. If the difference between two windows is greater than a predefined threshold, it considers that a concept drift happened, and the data from the first window is discarded~\cite{adwin}. Different from the other two methods, ADWIN has no warning level. Once a change occurs, the data that is out of the window is discarded and the remaining ones are used to retrain the classifier. Unsupervised drift detectors such as the ones proposed by Žliobaité et al. may be useful when delays are expected given that they do not rely on the real label of the samples, which need to be known by supervised methods, and most of the time in cybersecurity it does not happen in practice~\cite{5693384}. These unsupervised strategies consist in comparing different detection windows of fixed length using statistical tests over the data themselves, on the classifier output labels or its estimations (that may contain errors) to detect if both come from the same source. In addition, active learning may complement these unsupervised methods by requiring the labels of only a subset of the unlabeled samples, which could improve drift detection and overall classification performance.

\review{Some authors also created different classification models and strategies that deal with both concept drift and concept evolution.}
\review{Shao et al. proposed SyncStream, a classification model for evolving data streams that use prototype-based data representation, P-Tree data structure, and just a small set of both short and long-term samples based on error-drive representativeness learning (instead of using base classifiers or windows of data)~\cite{10.1145/2623330.2623609}.} 
\review{ZareMoodi et al. created a new supervised chunk-based method for novel class detection using ensemble learners, local patterns, and connected components of neighborhood graphs~\cite{ZAREMOODI2015234}. The same authors also proposed a new way to detect evolving concepts by optimizing an objective function using a fuzzy agglomerative clustering method~\cite{ZareMoodi19}.} 
\review{Hosseini et al. created SPASC (Semi-supervised Pool and Accuracy-based Stream Classification), an ensemble of classifiers where each classifier holds a specific concept, and new samples are used to add new classifiers to the ensemble or to update the existing ones according to their similarity to the concepts~\cite{Hosseini15}.} 
\review{Dehghan et al. proposed a method based on the ensemble to detect concept drift by monitoring the distribution of its error, training a new classifier on the new concept to keep the model updated~\cite{Dehghan2016ANC}.} 
\review{Ahmadi et al. created GraphPool, a classification framework that deals with recurrent concepts by looking at the correlation among features, using a statistical
multivariate likelihood test, and maintaining the transition among concepts via a first-order Markov chain~\cite{Ahmadi2017ModelingRC}.}
\review{Gomes et al. presented the Adaptive Random Forest (ARF) algorithm, an adaptation of the classical random forest algorithm with dynamic update methods to deal
with evolving data streams. The ARF also contains an adaptive strategy that uses a concept drift detector in each tree to track possible changes and to train new trees in the background~\cite{gomes2017adaptive}.}
\review{Finally, Siahroudi et al. proposed a method using multiple kernel
learning to detect novel classes in non-stationary data streams~\cite{SIAHROUDI2018187}. The authors do it by classifying each new instance by computing their distance to the previously known classes in the feature space and updating the model based on their true
labels.} 

We advocate for more collaboration between data stream learning and cybersecurity, given that the majority of cybersecurity works presented in this section do not use data stream approaches (including concept drift detectors), they both have a lot of practice problems in common and may benefit each other. For instance, data stream learning could benefit from real cybersecurity datasets that could be used to build real-world \ac{ML} security solutions, resulting in higher quality research that may also be useful in other \ac{ML} research fields. Finally, developing new drift detection algorithms is important to test their effectiveness in different cybersecurity scenarios and \ac{ML} models.

\subsection{Adversarial Attacks}

In most cybersecurity solutions that use \acl{ML}, models are prone to suffer adversarial attacks, where attackers modify their malicious vectors to somehow make them not being detected~\cite{10.1145/3375894.3375898}. \reviewdtrap{These techniques were proven effective in both malware and intrusion scenarios~\cite{9001114}, for instance.} We already mentioned this problem related to feature robustness in Section~\ref{subsec:adversarial_ml_features}, but \ac{ML} models are also subject to adversaries. These adversarial attacks may have several consequences such as allowing the execution of malicious software, poisoning an \ac{ML} model or drift detector if they use new unknown samples to update their definitions (without a ground truth from other sources), and producing, as a consequence, concept drift and evolution. Thus, when developing cybersecurity solutions using \ac{ML}, both features and models must be robust against adversaries.

Aside from using adversarial features, attackers may also directly attack \ac{ML} models. There are two types of attacks: white-box attacks, where the adversary has full access to the model, and black-box attacks, where the adversary has access only to the output produced by the model, without directly accessing it~\cite{DBLP:journals/corr/abs-1802-00420}. A good example of white-box attacks is gradient-based adversarial attacks, which consist in using the weights of a neural network to obtain perturbation vectors that, combined with an original instance, can generate an adversarial one that may be classified by the model as being from another class~\cite{goodfellow2014explaining}.  Many strategies use neural network weights to produce these perturbations~\cite{DBLP:journals/corr/abs-1802-00420}, which not only affects neural networks but a wide variety of models~\cite{goodfellow2014explaining}. Other simpler white-box attacks such as analyzing the model, for instance, the nodes of a decision tree or the support vectors used by an SVM, could be used to manually craft adversarial vectors by simply changing the original characteristics of a given sample in a way that it can affect its output label. In contrast, black-box attacks tend to be more challenging and real for adversaries, given that they usually do not have access to implementations of cybersecurity solutions or \ac{ML} models, i.e., they have no knowledge about which features and classifiers a given solution is using and usually only known which is the raw input and the output. Thus, black-box attacks rely on simply creating random perturbations and testing them in the input data~\cite{DBLP:journals/corr/abs-1905-07121}, changing characteristics from samples looking at instances from all classes~\cite{DBLP:journals/corr/abs-1802-00420}, or trying to mimic the original model by creating a local model trained with samples submitted to the original one, using the labels returned by it, and then analyzing or using this new model to create an adversarial sample~\cite{DBLP:journals/corr/PapernotMGJCS16}. 

In response to adversarial attacks, defenders may try different strategies to overcome them, searching for more robust models that make this task harder for adversaries. One response to these attacks is Generative Adversarial Networks (GANs), which are 2-part, coupled deep learning systems in which one part is trained to classify the inputs generated by the other. The two parts simultaneously try to maximize their performance, improving the generation of adversaries, that are used to defeat the classifier, and then used to improve their detection by training the classifier with them~\cite{goodfellow2014generative, DBLP:journals/corr/HuT17}. Another valid strategy is to create an algorithm that, given a malign sample, automatically generates adversarial samples, similar to data augmentation or oversampling techniques that insert benign characteristics into it, which are then used to train or update a model. This way, the model will learn not only the normal concept of a sample but also the concept of its adversaries' versions, which will make it more resistant to attacks~\cite{DBLP:journals/corr/PapernotMJFCS15, Grosse2017AdversarialEF}. 
\reviewdtrap{Instead of using the hard class labels, Apruzzese et al. propose using the probability labels to make random forest-based models more resilient to adversarial perturbations, achieving comparable or even superior results even in the absence of attacks~\cite{10.1109/TETCI.2019.2961157}}.

Some approaches also tried to fix limitations of already developed models, such as \texttt{MalConv}~\cite{roots_model1}, an end-to-end deep learning model, which takes as input raw bytes of a file to determine its maliciousness. \texttt{Non-Negative MalConv} proposes an improvement to \texttt{MalConv}, with an identical structure, but having only non-negative weights, which forces the model to look only for malicious evidence rather than looking for both malicious and benign ones, being less prone to adversaries that try to copy benign behavior~\cite{roots_model2}. Despite that, even \texttt{Non-Negative MalConv} has weaknesses that can be explored by attackers~\cite{10.1145/3375894.3375898}, which makes this topic an open problem to be solved by future research. We advocate for more work and competitions, such as the Machine Learning Security Evasion Competition (MLSEC)~\cite{mlsec20}, that encourage the implementation of new defense solutions that minimize the effects of adversarial attacks.

\subsection{Class imbalance}
\label{subsec:class_imabalance_model}

Class imbalance is a problem already mentioned in this work, but on the dataset side (Section~\ref{subsec:class_imbalance}). In this section, we are going to discuss the effects of class imbalance in the \ac{ML} model and present some possible mitigation techniques that rely on improving the learning process (cost-sensitive learning), using ensemble learning (algorithms that combine the results of a set of classifiers to make a decision) or anomaly detection (or one-class) models~\cite{kaur2019systematic, gomes19ds}. This way, when using cost-sensitive learning approaches, the generalization made by most algorithms, which makes minority classes ignored, is adapted to give each class the same importance, reducing the negative impact caused by class imbalance. Usually, cost-sensitive learning approaches increase the cost of incorrect predictions of minority classes, biasing the model in their favor and resulting in better overall classification results~\cite{kaur2019systematic}. Such techniques are not easy to implement in comparison to sampling methods presented in Section~\ref{subsec:class_imbalance} but tend to be much faster given that they just adapt the learning process, without generating any artificial data~\cite{gomes19ds}. 

In addition, ensemble learning methods that rely on bagging~\cite{10.1023/A:1018054314350} or boosting techniques (such as AdaBoost~\cite{10.1006/jcss.1997.1504}) present good results with imbalanced data~\cite{GalarClassImb2012}, which is one of the reasons that random forest performs well in many cybersecurity tasks with class imbalance problems, such as malware detection~\cite{nfs}. Bagging consists in training the classifiers from an ensemble with different subsets of the training dataset (with replacement), introducing diversity to the ensemble, and improving overall classification performance~\cite{10.1023/A:1018054314350,GalarClassImb2012}. The AdaBoost technique consists in training each classifier from the ensemble with the whole training dataset in iterations. After each iteration, the algorithm gives more importance to difficult samples, trying to correctly classify the samples that were incorrectly classified by giving them different weights, very similar to what cost-sensitive learning does, but without using a cost to update the weights~\cite{10.1006/jcss.1997.1504,GalarClassImb2012}.

Even though all the methods presented so far are valid strategies to handle imbalanced datasets, sometimes the distribution of classes is too different that it is not viable to use one of them, given that the majority of the data will be discarded (undersampling), poor data will be generated (oversampling), or the model will not be able to learn the concept of the minority class~\cite{gomes19ds}. In these cases, anomaly detection algorithms are strongly recommended, given that they are trained over the majority class only and the remaining ones (minority class) are considered anomalous instances~\cite{salehi2018survey, gomes19ds}. Two great examples of anomaly detection models are isolation forest~\cite{10.1145/2133360.2133363} and one-class SVM~\cite{10.5555/3009657.3009740}. Both of them try to fit the regions where the training data is most concentrated, creating a decision boundary that defines what is normal and what is an anomaly. 


Finally, when building an \ac{ML} solution that has imbalanced data, as well as testing several classifiers and feature extractors, it is also important to consider the approaches presented here for both the dataset and model sides. Also, it is possible to combine more than one method, for instance, generating a set of artificial data and using cost-sensitive learning strategies, which could increase classification performance in some cases. We strongly recommend that cybersecurity researchers include some of these strategies in their work, given that it is difficult to find solutions that actually consider class imbalance.

\subsection{Transfer Learning}
\label{subsec:transfer_learning}

Transfer learning is the process of learning a given task by transferring knowledge from a related task that has already been learned. It has shown to be very effective in many \ac{ML} applications~\cite{torrey2010transfer}, such as image classification~\cite{DBLP:journals/corr/abs-1902-07208, 10.1007/978-3-319-97982-3_16} and natural language processing problems~\cite{DBLP:journals/corr/abs-1801-06146, Radford2018ImprovingLU}. Recently, Microsoft and Intel researchers proposed the use of transfer learning from computer vision to static malware detection~\cite{DBLP:journals/corr/abs-1812-07606}, representing binaries as grayscale images and using inception-v1~\cite{DBLP:journals/corr/SzegedyLJSRAEVR14} as the base model to transfer knowledge~\cite{Chen2020}. The results presented by the authors show a recall of 87.05\%, with only a 0.1\% of false positive rate, indicating that transfer learning may help to improve malware classification without the need of searching for optimal hyperparameters and architectures, reducing the training time and the use of resources.

In addition, if the network used as the base model is robust, they probably contain robust feature extractors. Consequently, by using these feature extractors, the new model produced inherits their robustness, producing new solutions that are also robust to adversarial attacks, achieving high classification performances, without much data, and with no need to use a lot of resources as some adversarial training approaches~\cite{DBLP:journals/corr/abs-1905-08232}. 
At the same time that transfer learning might be an advantage, it may also be a problem according to the base model used because usually, these base models are publicly available, which means that any potential attackers might have access to them and produce an adversarial vector that might affect both models: the base and the new one~\cite{DBLP:journals/corr/abs-1904-04334}. Thus, it is important to consider the robustness of the base model when using it to transfer learning to produce a solution without security weaknesses. Finally, despite presenting promising results, the model proposed to detect malware by using transfer learning cited at the beginning of this subsection~\cite{Chen2020} may be affected by adversarial attacks, given that its base model is affected by them as already shown in the literature~\cite{goodfellow2014explaining, DBLP:journals/corr/CarliniW16a}. 


\subsection{Implementation}
\label{subsec:implementation}

Building a good \acl{ML} model is not the last challenge to deploying \ac{ML} approaches in practice. The implementation of these approaches might also be challenging~\cite{10.1145/3533378}. The existing frameworks, such as scikit-learn~\cite{scikit-learn} and Weka~\cite{weka}, usually rely on batch learning algorithms, which may not be useful in dynamic scenarios where new data are available all the time (as a stream), requiring the model to be updated frequently with them~\cite{gomes19ds}. In these cases, \ac{ML} implementations for streaming data, such as Scikit-Multiflow~\cite{skmultiflow}, \review{Massive Online Analysis (MOA)~\cite{moa}, River~\cite{2020river}}, and Spark~\cite{spark_streaming1, spark_streaming2} are highly recommended, once they provide \ac{ML} algorithms that could be easily used in real cybersecurity applications. Also, adversarial machine learning frameworks, such as CleverHans~\cite{papernot2018cleverhans} and SecML~\cite{melis2019secml}, are important to test and evaluate the security of \ac{ML} solutions proposed. Thus, contributing to streaming data and adversarial machine learning projects is as important as contributing to well-known \ac{ML} libraries, and we advocate for that to make all research closer to real-world applications. Note that we are not just talking specifically about contributing with new models, but also prepossessing and evaluating algorithms that may be designed only in batch learning, and could also be a good contribution to streaming learning libraries. We believe that more contributions to these projects would benefit both industry and academia with higher quality solutions and research, given the high number of research works using only batch learning algorithms nowadays, even in cybersecurity problems.

In addition, multi-language codebases may be a serious challenge when implementing a solution, once different components may be written in different languages, not being completely compatible, becoming incompatible with new releases, or being too slow according to the code implementation and language used~\cite{Schelter2018OnCI}. Thus, it is common to see \ac{ML} implementations being optimized by C and C++ under the hood, given that they are much faster and more efficient than Python and Java, for instance. Despite such optimizations being needed to make many solutions feasible in real-world solutions, they are not always performed given that (i) researchers create their solutions as prototypes that only simulate the real world, not requiring optimizations, and (ii) optimizations require knowledge about code optimization techniques that are very specific or may be limited to a given type of hardware, such as GPUs~\cite{Schelter2018OnCI}. Also, implementing data stream algorithms is a hard task, given that we need to execute the whole pipeline continuously: if any component of this pipeline fails, the whole system may fail~\cite{gomes19ds}.

Another challenge is to ensure a good performance for the proposed algorithms and models~\cite{10.1145/3352020.3352024}. A good performance is essential to deploy ML in the security context because most of the detection solutions operate in runtime to detect attacks as early as possible and slow models will result in a significant slowdown to the whole system operation. To overcome the performance barriers of software implementations, many security solutions opt to outsource the processing of ML algorithms to third-party components. A frequent approach in the security context is to propose hardware devices to perform critical tasks, among which is the processing of ML algorithms~\cite{Botacin2019RECOSOC}. Alternatively, security solutions might also outsource scanning procedures to the cloud. Many research works proposed cloud-based AVs~\cite{7831709,10.1145/2420950.2420983}, which have the potential to include ML-based scans among their detection capabilities and  streamline such checks to the market. We understand that these scenarios should be considered for the proposal of new ML-based detection solutions.
\section{Evaluation}
\label{sec:evaluation}

Knowing how to correctly evaluate an \ac{ML} system is essential to building a security solution, given that some evaluations may result in wrong conclusions that may backfire in security contexts, even when using traditional \ac{ML} evaluation best practices~\cite{226323, tesseract, nfs}. For instance, consider that a malicious threat detection model is evaluated using ten samples: eight of them are benign and two are malign. This model has an accuracy of 80\%. Is 80\% a good accuracy? Assuming that the model classifies correctly only the eight benign samples, it is not capable of identifying any malign sample, and yet it has an accuracy of 80\%, giving a false impression that the model works significantly well. Thus, it is important to take into account the metric used to produce high-quality systems that solve the problem proposed by a certain threat model. 

\subsection{Metrics}
\label{subsec:metrics}


To correctly evaluate a solution the right metrics need to be selected to provide significant insights that can present different perspectives of the problem according to its real needs, reflecting the real world~\cite{Gron2017}. One of the most used metrics by \ac{ML} solutions is accuracy, which consists in measuring the percentage of samples correctly classified by the model divided by the total number of samples seen by it (usually from the testing set)~\cite{FERRI200927}. The main problem with this metric is that it may provide wrong conclusions according to the distribution of the datasets used, as already shown by the malicious threat detection model example. Thus, if the dataset is imbalanced, accuracy is not recommended since it will give much more importance to the majority class, presenting, for instance, high values even if a minority class is completely ignored by the classifier~\cite{FERRI200927}.

An interesting way to evaluate model performance is by checking the confusion matrix, a matrix where each row represents the real label and each column represents a predicted class (or vice versa)~\cite{Gron2017}. By using this matrix, it is possible to check a lot of information about the model, for instance, which class is more difficult to classify or which ones are being confused the most. In addition, it is also possible to calculate false positives and false negatives, which leads us to recall, precision, and, consequently, f1score.
Recall measures the percentage of positive examples that are correctly classified, precision measures the percentage of examples classified as positive that are positive, and f1score is the harmonic mean of both of them~\cite{10.1145/1143844.1143874}. With these metrics, it is possible to calibrate the model according to the task being performed~\cite{nfs}. For a malware detector, for instance, it may be better to not detect malware than to block benign software (high precision), given that it would affect directly the user experience. Imagine that a user is using his computer and, when opening Microsoft Word, the classifier believes that it is malware and blocks its execution. Even detecting the majority of the malware, the classifier may turn the computer useless, as it would block a great part of the benign applications. In contrast, in more sensitive systems, one might want the opposite (high recall), blocking all the malign actions, even with some benign being ''sacrificed``.

Recently, some authors introduced metrics to evaluate the quality of the models produced. Jordaney et al. proposed Conformal Evaluator (CE), an evaluation framework that computes two metrics (algorithm confidence and credibility) to measure the quality of the produced \ac{ML} results, evaluating the robustness of the predictions made by the algorithm and their qualities~\cite{Jordaney2017}. Pendlebury et al. introduced another evaluation framework called \textsc{Tesseract}, which compares classifiers in a realistic setting by introducing a new metric, Area Under Time (AUT). This metric captures the impact of time decay on a classifier, which is not evaluated in many works, confirming that some of them are biased. Thus, we support the development of these kinds of work to better evaluate new \ac{ML} solutions considering a real-world scenario that allows the implementation to be used in practice.

\subsection{Comparing Apples to Orange}
\label{subsec:comparing_apples_to_orange}

It is not unusual for researchers to compare their evaluation results with other prior literature solutions that face the same problems (e.g., comparing the accuracy of two detection models). Such comparison, however, should be carefully performed to avoid misleading conclusions. 

The lack of standard publicly available repositories is responsible for lots of work using their own dataset when building a solution. These new solutions have their results usually compared with other works reported in the literature. Whereas comparing approaches seems to be straightforward, authors should care to perform fair evaluations, such as comparing studies leveraging the same datasets, avoiding presenting results deemed to outperform literature results but which do not achieve such performance in actual scenarios.

As an analogy, consider image classification problems whose objective is to identify objects represented in images (for instance, buildings, animals, locations, etc). These challenges often provide multiple datasets that are used as a baseline by many solutions. For instance, the CIFAR challenge~\cite{cifar} is composed of two datasets: CIFAR-100, which has one hundred classes of images, and CIFAR-10, which is a filtered version of CIFAR-100, containing just ten classes. Imagine two research work proposing distinct engineering solutions for image classification, one of them leveraging CIFAR-10 and the other leveraging CIFAR-100. Although one of the approaches presents a higher accuracy than the other, is it fair to say that this one is better than the other? No, because the task involved in classifying distinct classes is also distinct. The same reasoning is valid for any cybersecurity research involving ~\ac{ML}. Thus, authors should care to not perform comparisons involving distinct classes of applications, such as comparing, for instance, approaches involving Dynamic System Call Dependency Graphs, a computationally costly approach, with static feature extraction approaches. This is misleading because each type of work presents different natures and challenges. Finally, it is strongly recommended that researchers share their source codes with the community to make their work compatible with any other dataset (which in the majority of the works are not shared), allowing future researchers to compare different approaches in the same scenario.

\subsection{Delayed Labels Evaluation}
\label{subsec:delayed_labels}


One particularity of security data is that they usually do not have ground-truth labels available right after new data are collected, as already shown in Section~\ref{subsec:data_labeling}. Due to that, there is a gap between the data collection and their labeling process, which is not considered in many cybersecurity types of research that uses \ac{ML}, i.e., in many proposed solutions in the literature, the labels are available at the same time as the data, even in online learning solutions. Some of them ignore the ground-truth label and use the same labels that the \ac{ML} classifier predicts~\cite{Xu2019}, which may make the model subject to poisoning attacks and, consequently, decrease its detection rate as time goes by~\cite{taheri2019defending}. 

\begin{figure*}
     \centering
     \begin{subfigure}[t]{0.49\textwidth}
         \centering
         \includegraphics[width=\textwidth]{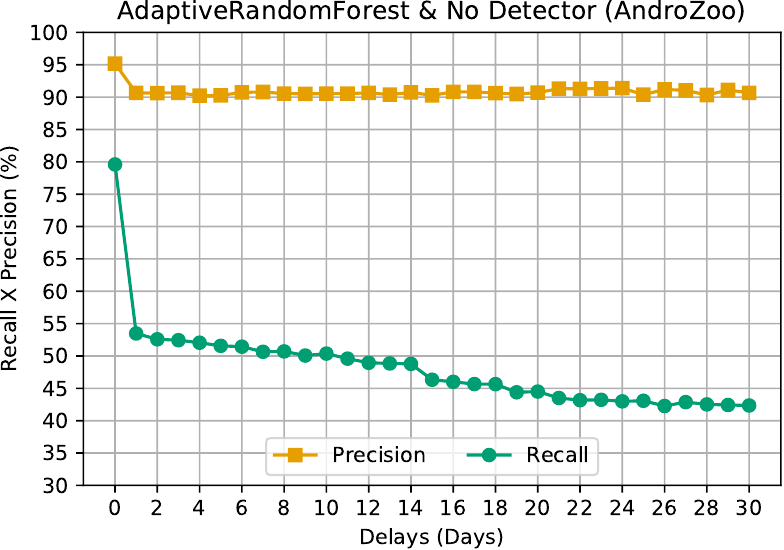}
         \caption{\textbf{AndroZoo dataset without drift detector.} Results for the adaptive random forest with no drift detector using AndroZoo.}
         \label{fig:delay_androzoo_01}
     \end{subfigure}
     \hfill
     \begin{subfigure}[t]{0.49\textwidth}
         \centering
         \includegraphics[width=\textwidth]{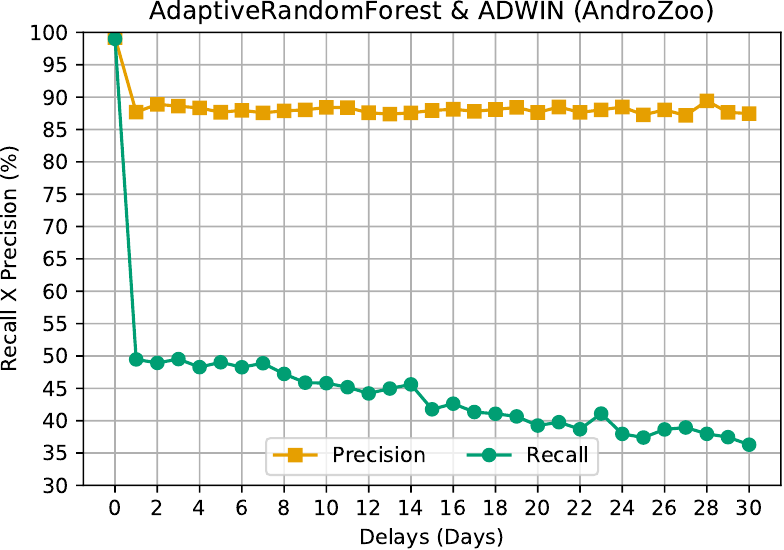}
         \caption{\textbf{AndroZoo dataset with drift detector.} Results for the adaptive random forest with ADWIN drift detector using AndroZoo.}
         \label{fig:delay_androzoo_02}
     \end{subfigure}

    \caption{\textbf{Delayed Labels Evaluations.} Models with drift detection may not perform as expected in real-world solutions, despite having the best performance in scenarios where delays are not considered.}
    \label{fig:delay_drebin}
\end{figure*}

Considering malware detection, the majority of the works use only a single snapshot of scanning results from platforms like VirusTotal, without considering a given delay before actually using the labels, which may vary from ten days to more than two years~\cite{251586}. A recent study from Botacin et al. analyzed the labels provided by VirusTotal for 30 consecutive days from two distinct representative malware datasets (from Windows, Linux, and Android) and showed that solutions took about 19 days to detect the majority of the threats~\cite{BOTACIN2020101859}. To study the impact of these delayed labels in malware detectors using \ac{ML}, we simulated a scenario using the AndroZoo dataset~\cite{Allix:2016:ACM:2901739.2903508} and online \ac{ML} techniques with and without drift detectors by providing the labels of each sample $N$ days after they are available to further update the decision model (Adaptive Random Forest~\cite{gomes2017adaptive} trained using TF-IDF feature extractor). The results of this experiment are shown in Fig.~\ref{fig:delay_androzoo_01} and Fig.~\ref{fig:delay_androzoo_02}. We can note that when not considering a delay, using a drift detector improves the detection rate a lot. However, after one day of delay, this is not true anymore: the model that does not consider concept drift performs better overall, despite both being very affected by this problem, dropping to half of the original precision in the case of Adaptive Random Forest with ADWIN. These results indicate that: (i) in scenarios where delayed labels exist, \ac{ML} models do not perform the way they are evaluated without these conditions, and (ii) models that make use of drift detectors, such as ADWIN, present lower detection rates in comparison to those that do not use them, since the concept being learned by the model is outdated when a drift is detected, increasing false negatives.

Finally, we advocate for drift detection strategies that consider the delay of labels and mitigation techniques to overcome this problem, such as active learning or approaches to provide real labels with less delay, to produce better solutions~\cite{10.1145/2674026.2674028}.

\subsection{Online vs Offline Experiments} 
\label{subsec:online_vs_offline_experiments}


We previously advocated for real-world considerations when developing an \ac{ML} model (Section~\ref{sec:model}). We also advocate for real-world considerations when evaluating the developed solutions. Each evaluation should consider the scenario in which it will be applied. A major implication of this requirement is that online and offline solutions must be evaluated using distinct criteria. Offline solutions are often leveraged by security companies in their internal analysis (e.g., an AV company investigating whether a given payload is malicious or not to develop a signature for it that will be distributed to the customers in the future). Online solutions, in turn, are leveraged by the endpoints installed in the user machines (e.g., a real-time AV monitoring a process behavior, a packet filter inspecting the network traffic, and so on). Due to their distinct nature, offline solutions are much more flexible than online solutions. Whereas offline solutions can rely on complex models that are run on large clusters, online solutions must be fast enough to be processed in real time without adding a significant performance overhead to the system. Moreover, offline solutions can collect huge amounts of data during a sandboxed execution and thus input them into models relying on a large moving window for classification. Online solutions, in turn, operate in memory-limited environments and aim to detect the threat as soon as possible. Thus, they cannot rely on large moving windows. These differences must be considered when evaluating the models to avoid common pitfalls, such as applying the same criteria for both applications. Due to their distinct nature, online solutions are expected to present more false negatives than offline solutions, as they have to make fast decisions about the threat. On the other hand, offline solutions will detect more samples, as they have more data to decide, but the detection is likely to happen later than in an online detector, only after multiple windows (e.g., when malware already infected the system and/or when a remote payload already exploited a vulnerable application). For a more complete security treatment, modern security solutions should consider hybrid models that employ multiple classifiers, each one with a distinct window size~\cite{9061034}. 

\section{Discussion: Understanding ML}
\label{sec:discussion}

Once we have presented all the steps required to apply \ac{ML} to security problems, we now discuss how this application should be understood, its limitations, and existing open gaps.

\subsection{A ML model is a type of signature}

A frequent claim of most products and research work leveraging \ac{ML} as part of their operation, mainly antivirus and intrusion detection systems, is that this use makes their approaches more flexible and precise than the so-far applied signature schemes~\cite{sigml}. This is somehow a pitfall, as, in the last instance, an \ac{ML} model is just a weighted signature of boolean values. One can even convert an \ac{ML} model to typical YARA signatures~\cite{mltoyara}, as deployed by many security solutions. Although the weights can be adjusted to add some flexibility to an \ac{ML} model, the model itself cannot be automatically extended beyond the initially considered boolean values without additional processing (re-training or feature re-extraction), which is an already existing drawback for signature schemes. In this sense, the adversarial attacks against the \ac{ML} models can see as analogous to the polymorphic attacks against typical signature schemes~\cite{10.1145/2645791.2645857}: whereas a typical polymorphic threat attempt to directly change its features to pass the checks (weight=1), a modern attack against \ac{ML} models indirectly attempts to do so by presenting a relative feature frequency higher or lower than the considered in the \ac{ML} model.

\subsection{There is no such thing as a 0-day detector}

It is usual for many \ac{ML}-based detector proposals to state that their approach is resistant to 0-day attacks (i.e., attacks leveraging so-far unknown threats and/or payloads) because they rely on an \ac{ML} model~\cite{9006514,8473321}. This is also somehow a pitfall, and we credit this as the poor understanding that \ac{ML} models are a type of signature, as previously presented. In fact, the ability of a model to detect a new threat depends on the definition of what is new. If a new sample is understood as any sample that an attacker can create, certainly \ac{ML} models will be able to detect many 0-days, as many of these new payloads will be variations of previously known threats, a scenario for which weighted signatures generalize well. However, if we consider as new something unknown to the model in all senses (e.g., a payload having a feature that is not present in the model), no \ac{ML} model will be able to detect it as malicious, thus nature of the presented problem of concept drift and evolution. Therefore, \ac{ML} models should not be understood as simply more resistant detectors than typical signature schemes, but as a type of signature that can be more generalized.

\subsection{Security \& Explainable Learning}

Most
research works put a big focus on achieving good detection rates, but not all of them put the same effort into discussing and explaining the results~\cite{arp2014}, which would be essential to allow security incident response teams (CSIRTs) to fill the identified security breaches. The scenario is even more complicated when deep learning approaches are considered, as in most cases there is no clear explanation for the model operation~\cite{10.1145/3366424.3383110}. 
A model to detect cybersecurity threats can yield valuable insights besides its predictions since based on the explanation provided by the model, the security around the monitored object can be improved in future versions. 
As an example, a model able to explain that some threats were identified due to the exploitation of a given vulnerability might trigger patching procedures to fix the associated applications. 
Explainability has different levels of relevance according to the domain; we argue that for several cybersecurity tasks, they are essential to make it possible to apply countermeasures to security threats. 




\subsection{The arms race will always exist}
\label{subsec:arms_race}

\begin{figure*}[!htpb]
    \centering
    \includegraphics[width=.775\textwidth]{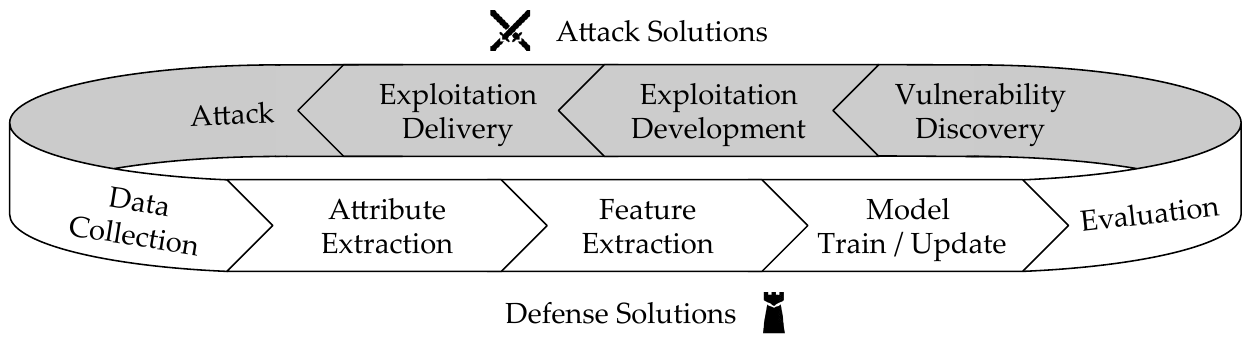}
\caption{\textbf{Attack VS Defense Solutions.} The arms race created by both generates a never-ending cycle.}
    \label{fig:attack_vs_defense_solutions}
\end{figure*}

The arms race between attack and defense solutions is a cycle that will always exist, requiring that both sides keep investing in new approaches to overcome their adversaries. Thus, new approaches from one the sides result in a reaction from another one, as shown in Fig.~\ref{fig:attack_vs_defense_solutions}, where defense solutions follow the steps mentioned in this work to build \ac{ML} defense solutions and attackers follow the steps required to create an attack: they first find the weakness of the defense solutions, use this weakness to develop an exploit, which is then delivered to be executed, producing new data for \ac{ML} models to be trained, restarting the cycle~\cite{10.1145/3199674}. Finally, we advocate that defense solutions try to reduce the gap present in this cycle (the development of new attacks and the generation of solutions for them) by following the recommendations of this work \review{and other works in the literature that make use of robustness verification to prevent attacks against adversaries~\cite{li2020sok}}.

\review{To help the development of future ML solutions for cybersecurity, we created a checklist\footnote{The draft version of the checklist is available at 
\blind.} that is an adaptation based on the challenges, pitfalls, and problems reported in this work. Thus, anyone developing or reviewing a solution can fill the questions on this checklist and get feedback reporting what could be improved or corrected according to our findings.}

\subsection{The future of ML for cybersecurity}

ML for cybersecurity has become so popular that is hard to imagine a future without it. However, this does not mean that there is no room for improvement. As an immediate task, researchers will put their effort to mitigate adversarial attacks, which will enable even more applications to migrate ML solutions. A massive migration will require not only robust security guarantees but also privacy ones, which is likely to be achieved via merging \ac{ML} and cryptography. The field is experiencing the rise of \ac{ML} classifiers based on homomorphic cryptography~\cite{8241854}, which allows data to be classified without decryption. We believe that this type of approach has the potential to upstream many \ac{ML}-powered solutions, such as cloud-based AVs, which will be able to scan files in the cloud while preserving the privacy of the file's owners.

\section{Conclusion}
\label{sec:conclusion}

In this paper, we introduced a set of problems and challenges that have been observed (too often) in \ac{ML} techniques applied to diverse cybersecurity solutions. We presented practical examples of cybersecurity scenarios in which \ac{ML} might either be incorrectly applied or contain blind spots (important details that were not considered, discussed, or observed). When possible, we showed techniques to address those common issues. 
To make a step toward that, in this paper, we summarized and provided insights on the following main points: data collection issues (data leakage, data labeling, class imbalance, and dataset size definition); modeling (different attribute and feature extraction methods, adaptation to changes, i.e. concept drift/evolution, adversarial features and model attacks, one-class models, cost-sensitive, ensemble learning, transfer learning, and implementation challenges); and evaluation concerns (adequate use of metrics, thorough comparison of previous/existing solutions, delayed labels, and online vs. offline experiments).


Finally, our main recommendations to improve \ac{ML} in security are the following:

\begin{itemize}
    \item \textbf{Stop looking only at metrics, and start looking at effects:} many of the challenges presented in this work remain as open research problems, which would benefit both academia and industry if properly solved. In addition, their mitigation would foster the use of ML for cybersecurity problems, and improve the cybersecurity field in the same way ML advanced other research fields. Community players (security and machine learning) have to take into consideration the plenty of peculiarities associated with cybersecurity data and its sources.
    \item \textbf{Commit yourself to the real world:} we advocate that future research works always keep the motto “machine learning that matters”~\cite{DBLP:journals/corr/abs-1206-4656} in mind when developing new solutions. If your work is losing connection to problems of the real world, science, and society, we have a problem!
    \item \textbf{Check your work:} our recommendation is to carefully observe all the items, and to consider each of them during the design and implementation of ML models for cybersecurity. To encourage that, we presented a checklist (draft version available at 
    \blind;
    validation tests ongoing) to serve as a reminder and prevent researchers and practitioners from committing these common mistakes or at least being aware of their existence.
\end{itemize}

\bibliographystyle{ACM-Reference-Format}
\bibliography{references}

\end{document}